# Reward Sharing Schemes for Stake Pools

Lars Brünjes[*]    Aggelos Kiayias[†]    Elias Koutsoupias[‡]    Aikaterini-Panagiota Stouka[†]

Saturday 6th June, 2020

**Abstract**

We introduce and study reward sharing schemes (RSS) that promote the fair formation of *stake pools* in collaborative projects that involve a large number of stakeholders such as the maintenance of a proof-of-stake (PoS) blockchain. Our mechanisms are parameterized by a target value for the desired number of pools. We show that by properly incentivizing participants, the desired number of stake pools is a Nash equilibrium arising from rational play. Our equilibria also exhibit an efficiency / security tradeoff via a parameter that calibrates between including pools with the smallest cost and providing protection against Sybil attacks, the setting where a single stakeholder creates a large number of pools in the hopes to dominate the collaborative project. We then describe how RSS can be deployed in the PoS setting, mitigating a number of potential deployment attacks and protocol deviations that include censoring transactions, performing Sybil attacks with the objective to control the majority of stake, lying about the actual cost and others. Finally, we experimentally demonstrate fast convergence to equilibria in dynamic environments where players react to each other's strategic moves over an indefinite period of interactive play. We also show how simple reward sharing schemes that are seemingly more "fair", perhaps counterintuitively, converge to centralized equilibria.

## 1 Introduction

One of the main open questions in blockchain systems research is developing reward mechanisms that incentivize honest protocol execution and decentralization. Bitcoin, the dominant example of proof-of-work blockchains, has been criticized for its susceptibility to protocol deviation attacks (e.g., selfish-mining [16] and mining games [24]), its tendency to centralise via the creation of mining pools [1, 43, 28, 21], and its high-energy expenditure. To address mainly the latter problem, many proof-of-stake (PoS) [30, 25, 13, 6] blockchains have been proposed. Despite progress in the understanding of the security properties of PoS blockchains, designing a robust incentive mechanism that promotes decentralization remains open.

We can abstract the problem that is to be solved as follows. Consider a society of agents that have stake in a joint effort that is recorded in a ledger and want to run a *collaborative project* (which might be maintaining the ledger itself). Stakeholders actively engaged in the project will incur operational costs (potentially different across the stakeholder population) and hence the project should provide some rewards to offset these costs. The stakeholders have the option to actively participate in maintenance or abstain from it. We will assume that the project can draw funds from a reward pool enabling, potentially at regular intervals, to distribute in some way a reward $R$ to the stakeholders. In the PoS setting, the reward pool can be facilitated either via the creation of new cryptocurrency, the collection of transaction fees, or a combination thereof. A viable solution would thus be in the form of a *reward sharing scheme* which will take as input the current snapshot of the collaborative project and distribute the rewards $R$ to all stakeholders. The aim is that, after potentially multiple iterations of reward sharing, there are still agents, who incentivized by the rewards, are engaged in maintenance (for if not, the project should be considered dead). Beyond being viable, a solution also needs to possess additional desirable characteristics, e.g., it is decentralised in the sense that a sufficient number of *distinct* stakeholders are active in the project.

---

[*]IOHK, `lars.bruenjes@iohk.io`
[†]University of Edinburgh & IOHK, `Aggelos.Kiayias,A.stouka@ed.ac.uk`
[‡]University of Oxford, `elias.koutsoupias@cs.ox.ac.uk`

There are three dominant approaches that have been considered in the PoS context. In the "direct democracy" approach, every stakeholder participates proportionally to their stake, which has downside that the operational costs can be so high that they discourage participation from small stakeholders resulting in so-called "whales" completely dominating the system or, in the worst-case, having operations stopping altogether. In the "jury" approach, followed by PoS systems like [30, 12], a random subset of $k$ stakeholders is elected at various intervals to carry out the task, which has the downside that either the jury tenure is short and most of the nodes need to be either constantly operationally ready without necessarily doing anything, or the jury tenure is long (or predictable way ahead of time) and then the risk of someone subverting the project by paying the elected nodes with small stake is high. Finally, in the "representative democracy" approach, broadly followed by [25, 11, 27], the stakeholders can empower other stakeholders to represent them in project maintenance and subsequently share the rewards. Given that empowering is performed via stake as recorded in the ledger, representatives can be thought to form "*stake pools*" in analogy to the mining pools of Bitcoin. The focus of this work is to develop reward mechanisms and analyze them game theoretically for this third approach.

**Our Results.** In our setting there are $n$ agents or players with stakes $s = (s_1, \ldots, s_n)$ and a private vector of costs for running a stake pool $c = (c_1, \ldots, c_n)$ for each one of players, should any of them choose to do so. Note that without loss of generality we assume $s_i, c_i \in (0, 1)$ for all $i$ and $\sum_i s_i \leq 1$. The stake is publicly recorded in some way but *without necessarily identifying how much stake belongs to each player, the player identities, or even their number $n$*. The cost stems from the inherent task of maintenance the players are supposed to perform if they setup a pool; in the PoS setting which is our primary focus that would be the cost of setting up a server that receives, organizes and verifies transactions to be recorded in the ledger. Each player mainly decides whether to participate directly or delegate its stake to another stakeholder to act on their behalf — or even split its stake into multiple such activities (see below about "Sybil behavior"). Delegation creates *pools* of stakeholders, where each pool consists of its leader who participates directly and its members that delegate their stake to the pool. The game is determined by the reward scheme that determines the way by which the total reward $R$ is distributed to the pools and how individual pool rewards are distributed to the pool members. Looking ahead, we will focus on the class of reward schemes that allocate reward $r(\sigma, \lambda)$ to a pool of total stake $\sigma$ and allocated pool leader stake $\lambda$; we call $r$ the *reward function*. The other component of a reward scheme determines how the pool reward $r(\sigma, \lambda)$ is distributed to the pool leader and pool members. It makes sense that the reward for the pool leader is different from the reward for pool members to compensate the pool leader for the cost it incurs by contributing to the collaborative project as well as to incentivize them to take the initiative to form a pool. We focus on reward schemes that distribute the pool reward as follows: the pool leader gets an amount to cover its cost of running the project as well as a fraction $m_j$ of the remaining amount which we call its (profit) *margin*. The remaining amount is distributed among the pool members, including the pool leader, *proportionally* to the stake that they contributed to the pool. In our analysis we will take advantage of automatic enforcement of our reward scheme, as e.g., this can be guaranteed by a smart contract built-in the underlying ledger.

Given a reward sharing scheme that belongs to the above class, the players will pick their strategy that determines whether they will run a pool or not and whether they will allocate some or all of their stake to pools created by other players. Natural questions about these games are: Do they have pure equilibria? Do they possess desirable properties such as decentralisation? Do the best-response dynamics converge fast to them?

An important and interesting observation here is that the standard notion of utility and Nash equilibrium for this game fails to capture what we intuitively expect to happen. The reason is that at a Nash equilibrium the players do not have to take into account the impact their selection will make on the moves of the other players. In particular, all Nash equilibria (if they exist) will have margins $m_j = 1$ for a simple reason: once the other players select their strategies and in particular the allocation of their stake, the best response of a pool leader is to increase its margin as much as possible. Similar situations occur in other games, such as the Cournot competition [19]. The appropriate framework for such games is to consider *non-myopic* utilities, i.e., consider equilibria in a setting where utility is defined in a non-myopic fashion, accounting for the effects that a certain move of a player will incur anticipating a strategic response by the other players.

Our main result is the introduction and analysis of a novel reward sharing scheme that is parameterized by (1) the desired number of pools $k$, and (2) a Sybil resilience parameter $\alpha$. The two parameters can be selected to fine-tune two desirable properties of the resulting configuration. The primary property is *decentralisation and fairness*, which is captured by the creation of $k$ pools of roughly the same size $1/k$. The secondary property we are interested in is *Sybil resilience*, which is captured by being able to influence the equilibrium configuration so that it takes the parties' stake into account. Our mechanism is described in the following definition.

**Definition 1** (A Sybil-resilient cap-and-margin reward scheme). Given a target number of pools $k \in \mathbb{N}$, and a Sybil resilience parameter $\alpha \in [0, \infty)$, the reward function $r(\sigma, \lambda)$ of a pool with total stake $\sigma$, out of which $\lambda$ stake belongs to the pool leader, is proportional to $\sigma' + \alpha' \lambda$, i.e.,

$$r(\sigma, \lambda) \sim \sigma' + \alpha' \lambda, \tag{1}$$

where $\sigma' = \min\{\sigma, \beta\}$, $\beta = 1/k$, and $\alpha' = \alpha \frac{\sigma' - \lambda \cdot (1 - \sigma'/\beta)}{\beta}$. The proportionality factor is selected so that the sum of rewards does not exceed the available funds.

If the primary aim of the reward scheme, i.e., to have pools of size $\sigma = \beta = 1/k$, is achieved, then $\alpha' = \alpha$ and the expression in the reward function simplifies to $r(\sigma, \lambda) = \sigma + \alpha \lambda$, that is, a linear combination of the pool stake and the stake of the pool leader. The expression in $r(\sigma, \lambda)$ for pool size $\sigma \neq \beta$ has been selected to get a Nash equilibrium with the desired properties. Note also that when a pool has stake $\sigma \geq \beta$, the additional stake above $\beta$ is essentially ignored. We will call such a pool *saturated*.

Our main theorem about this reward sharing scheme is the following.

**Theorem 1** (Informal statement). *There exists a Nash equilibrium for the reward scheme of Definition 1 that satisfies*

- *exactly $k$ pools are created, each of size equal to $1/k$,*

- *the pool leaders are the players with the highest value of*

$$P(s_i, c_i) = r(\beta, s_i) \cdot \frac{R}{1 + \alpha} - c_i, \tag{2}$$

*where $s_i$ and $c_i$ are the stake and cost of player $i$, and $R$ is the total reward distributed to the players, and*

- *the players have no incentive to lie about their cost $c_i$.*

The quantity $P(s_i, c_i)$ in (2) is the *potential profit* of stakeholder $i$ when this player creates a pool using their whole stake $s_i$ and the pool attracts total stake $\beta$.

It follows immediately from the above theorem that we obtain an equilibrium that achieves the primary decentralization and fairness objective. Regarding Sybil resilience, observe that the potential of the players is controlled by the parameter $\alpha$. When $\alpha = 0$, the pool leaders are the players with the smallest cost (resulting in the most cost-effective equilibrium) while as $\alpha$ grows, the stake backing up the pools starts to become more and more relevant in the equilibrium configuration, with the extreme case when $\alpha \to \infty$ and the costs of all players are roughly equal when the stakepools will be managed by backing up each pool with the largest amount of stake possible. We illustrate how we can facilitate Sybil resilience by calibrating the $\alpha$ parameter in the sense that any Sybil behaving player at the equilibrium has to invest resources linear in the number of identities (i.e., stake-pools in our setting) that they register, arguably the best one can hope for in the anonymous setting we operate. We note that although the above reward function may at first appear rather complicated, there is a strong justification behind it (cf. Section 4).

*Non-myopic utility and dynamics.* We also tackle the question of whether the equilibrium guaranteed by our theoretical analysis is effectively reachable when players are engaged in the game. We consider non-myopic dynamics with players applying a natural best-response strategy to each other's moves in succession. Specifically, the players compute the desirability of each announced pool, which is the answer to the following question of the players: "if I allocate a small stake $x$ to pool $j$, how much do I expect to gain?". In other words, the desirability is the *marginal reward* of pool $j$ provided that it will become a successful pool and obtain stake $\beta$. A non-myopic player then assumes that each of the $k$ most desirable pools will increase in size to become saturated and the remaining pools will end up with the stake of their pool leader, and allocates its stake accordingly. The player is non-myopic as they judge pools by their potential to issue profits, not their current membership size which potentially might be quite small especially at the beginning of the game. For pool leaders the situation is similar, but they have also to compute their margin. To do so, they calculate the maximum possible margin that still allows them to be one of the $k$ most desirable pools. The question then is whether these dynamics converge? how fast? and to which equilibrium? We provide experimental evidence that under reasonable assumptions of the stake distribution (for example, Pareto distribution) and of the cost distribution (for example, uniform distribution in an interval), the dynamics converge quickly to our Nash equilibrium that has $k$ saturated pools and the characteristic that all pools are formed by the players that are ranked best according to their potential profit as predicted by the theoretical analysis.

*Equilibria and incentive compatibility.* Our reward sharing scheme has a Nash equilibrium in which the reward is distributed fairly among all stakeholders, except for pool leaders that get an additional gain (Proposition 2). A nice property of this additional gain is that, all else being equal, it increases by at most $\delta x$ whenever the pool leader's cost decreases by $\delta x$. This means that our reward sharing scheme is incentive compatible: no player will benefit by lying about its cost.

*Deployment considerations in the PoS setting.* We provide a comprehensive list of potential attacks and deviations as well as how they are mitigated in a deployment of our RSS in the setting of a PoS protocol such as [25]. These include "rich get richer" considerations censorship[1] and Sybil attacks, as well as how to deal with underperforming pool leaders that fail to meet their obligations in terms of maintaining the service.

**Related work.** A number of previous works considered the incentives of mining pools in the setting of PoW-based cryptocurrencies (as opposed to PoS-based ones) such as Bitcoin [37, 38, 15, 41]. The main differences between mining pools in Bitcoin and stake pools in our setting are that (i) in Bitcoin all pool members perform mining and hence incur costs, while in PoS setting, only the pool leader runs the underlying protocol and incurs a cost while delegators have no cost, (ii) in Bitcoin each pool leader can choose a different way to reward pool members/miners while in our setting we prescribe a specific way for rewards to be shared between pool members. Regarding centralization, Arnosti and Weinberg, [1], have established that some level of centralisation takes place in Bitcoin in settings where differences in electricity costs are present between the miners. Also according to [28] in a setting where each unit of resource has a different value depending on the distribution of the resources among the players, miners have incentives to create coalitions. These results are inline with our (even more centralised) negative result on fair RSS's for the PoS setting, cf. Section 2.2. Another aspect we do not explore here, is the instability of such protocols when the rewards come mostly from transaction fees; this was explored in [7, 43].

---

[1] A censorship attack happens when the current pool leaders block new pool registrations.

With respect to PoS blockchain systems, a different and notable approach to stake pools is to use the stake as voting power to elect a number of representatives, all of equal power, as in delegated PoS (DPoS) [27]; for example, the cryptocurrency EOS [22] has 21 representatives (called block producers). This type of scheme differs from ours in that (i) the incentives of voters are not taken into account thus issues of low voter participation are not addressed, (ii) elected representatives, despite getting equal power, are rewarded according to votes received; this inconsistency between representation and power may result in a relatively small fraction of stake controlling the system (e.g., at some point, controlling EOS delegates representing just 2.2% of stakeholders was sufficient to halt the system,[5] which ideally could withstand a ratio less than 1/3), (iii) it may leave a large fraction of stakeholders without representation (e.g., in EOS, at some point, only 8% of total stake is represented by the 21 leading delegates[2]). Yet another alternative to stake pools is that of Casper [6], where players can propose themselves as "validators" committing some of their stake as collateral. The committed stake can be "slashed" in case of a proven protocol deviation. This type of scheme differs from ours in that (i) stakeholders wishing to abstain from protocol maintenance operations have no prescribed way of contributing to the mechanism (as in the case of voting in DPoS or joining a stake pool in our setting), (ii) a small fraction of stake may end up controlling the system while at the same time leaving a lot of stake decoupled from the protocol operation; this is because substantial barriers may be imposed in becoming a validator (e.g., in the EIP proposal for Casper[3] it is suggested that 1500 ETH will be the minimum deposit, which, at the time of writing is more than $370K$); this can make it infeasible for many parties to engage directly; on the other hand reducing this threshold drastically may make the entry barrier too low and hence still allow a small amount of stake to control the system. As a separate point, it is worth noting that for both the above approaches there is no known game theoretic analysis that establishes a similar result to the one presented herein, i.e., that the mechanism can provably lead to a Nash equilibrium with desirable decentralisation characteristics that include a high number of protocol actors and Sybil attack resilience. The compounding of wealth in PoS cryptocurrencies was studied in [17] where a new notion denoted by "equitability" is introduced to measure how much players can increase their initial fraction of stake. Also they prove that a "geometric" reward function is the best choice for optimizing equitability under certain assumptions; we remark that it is a folklore belief that PoS systems are inherently less equitable than ones based on PoW, however this belief seems to be unfounded, cf. [23]. With respect to equitability we show that by calibrating our Sybil resilience parameter to be small our system becomes "equitable" in the sense of providing similar rewards to stake pool leaders independently of their wealth.

From a game-theoretic perspective, our setting has certain similarities to cooperative game theory in which coalitions of players have a value. In our setting the players have weights (stake) and they are allowed to split it into various coalitions (pools). Our objective is to have a given number of equal-weight coalitions, which contrasts with the typical question in cooperative game theory on how the values of the coalitions are distributed (e.g., core or Shapley value) in such a way that the grand coalition is stable [33]. Actually, the games that we study are variants of congestion games with rewards on a network of parallel links, one for every potential pool. The reward on each link is determined by the reward function, which essentially determines an atomic splittable congestion game. But unlike simple atomic splittable congestion games [31], our games have different reward for pool leaders and for pool members. There are two main research directions for such games: whether they have unique equilibria and how to efficiently compute them [4]. Regarding the question of unique inner equilibria the most relevant paper to our inner game is [32] (but see also [36,3]) which shows that under general continuity and convexity assumptions, games on parallel links have unique equilibria. However, the conditions on convexity do not meet our design objectives and they do not seem to be useful in our setting.

Our work is related to two aspects of delegation games, which are games that address the benefits and other strategic considerations for players delegating to someone else to play a game on their behalf, such as owners of firms hiring CEO's to run a company. The first aspect is somewhat superficially related to this work in pool formation the pool members delegate their power to pool leaders. The second aspect which is much more relevant to our approach is that delegation changes the utility of the players (for example, by considering "credible threats" [39, 40]) or creates a two-stage game [44, 18, 42]. A typical two-stage delegation game is non-myopic Cournot competition [19] in which in the outer game the firms (players) decide whether to be profit-maximizers or revenue-maximizers, while in the inner game they play a simple Cournot competition [29]. Unlike our case, the inner Cournot competition has a simple unique equilibrium which defines a simple two-stage game.

---

[2]Statistics extracted from http://eos.dapptools.info/#/block-producers on July 27th, 2018.
[3]See https://eips.ethereum.org/EIPS/eip-1011.

Another research area that is relevant to this work is mechanism design, because participants may have an incentive not to reveal their true parameters, e.g., the cost for running a pool [31, 45].

In the proof of work setting, [20] considers reward sharing rules for proof-of-work systems under the assumption of discounted expected utility and identifies schemes that achieve fairness. Furthermore, an axiomatic approach to reward schemes of proof-of-work systems is taken in [9] in order to study fairness, symmetry, budget balancing and other properties. Unlike our work that considers *incentives for pool formation* with desirable properties, these two papers study intrinsic properties of the system *given* an existing pool formation.

Finally, after the first version of the present paper was made public (on the arXiv repository, cf. [5]), another work, [26], studied a parameterized notion of decentralization, where, in an ideal system, all participants should exert the *same power* in running the system, independently of their stake. This is a significantly more demanding notion of decentralization than the one considered here, where in an ideal system, participants *exert power proportional to their stake*. It is argued in [26] that in order for a system to achieve full decentralization, there must exist a strictly positive Sybil cost, that is, the cost of running two or more nodes should be higher when the nodes belong to the same entity than to multiple entities. Clearly in systems with anonymous users, Sybil costs cannot be positive and such concept of decentralization is impossible.

The construction we present has also been implemented and deployed on the Cardano incentivised testnet[4] in tandem with the Ouroboros protocol [25] with an outcome in line to our theoretical and experimental analysis. The parameters selected for the testnet were $k = 100, \alpha = 0$. At the time of this writing, more than 900 pools have been created with the first 100 pools controlling approximately 70% of the stake.

## 2 Reward Sharing Schemes

### 2.1 Model and Definitions

There are $n$ stakeholders (aka players) with stakes $s = (s_1, \ldots, s_n)$ such that $\sum_{i=1}^{n} s_i = 1$ and costs $c = (c_1, \ldots, c_n)$ (all assumed non-zero real values). The value $s_i$ represents the $i$-th player's stake in the *collaborative project* (which is e.g., maintaining the blockchain), while the value $c_i$ represents the $i$-th player's cost, should he decide to be active in the project's maintenance. The players want to engage in the collaborative project and *each player decides whether to participate directly by activating its pool or delegate his stake to other stakeholders.* The total stake that is delegated to an active stakeholder $j$ (note that the sum of all players' stakes is 1 so with the term "stake" we mean relative stake) forms a *stakepool*; we will call such a pool $\pi_j$, indexed by its *pool leader* $j$, and we will denote by $\sigma_j$ the total stake delegated to this pool by all players, including the pool-leader $j$. We will use $a_{i,j}$ to denote the stake that player $i$ allocates to pool $\pi_j$. The pools participate in the collaborative project through their leaders and this participation incurs cost $c_j$ for pool leader $j$. This cost is fixed for each player and does not depend on the size of the pool. To incentivize the stakeholders and pool leaders to form pools and work for the collaborative project, we introduce a *reward scheme*. We assume that there is a *fixed reward R* to be distributed among all pools. A *reward scheme* determines the way by which

---
[4]See, https://staking.cardano.org

*Notation*

- $n \in \mathbb{N}$, number of players.
- $k \in \mathbb{N}, k < n$, the desired number of pools.
- $R \in \mathbb{R}$, total reward.
- $s_i \in (0,1)$, stake of player $i$. It holds $\sum_{i=1}^{n} s_i = 1$.
- $c_i \in (0, R)$, cost of player $i$ to form a pool $\pi_i$.
- $m_i \in [0,1]$, margin of pool $\pi_i$.
- $\lambda_i \in (0,1)$, stake that player $i$ will commit if he activates his own pool $\pi_i$.
- $\vec{a}_i = (a_{i,1}, \ldots a_{i,n})$, allocation of player's $i$ stake. $\sum_j a_{i,j} \le s_i$.
- $\sigma_j$, stake of pool $\pi_j$: $\sigma_j = \sum_{i=1}^{n} a_{i,j}$. We denote the vector of pool stakes by $\vec{\sigma} = (\sigma_1, \ldots, \sigma_n)$. Pools can have zero stake.
- $r(\sigma, \lambda)$, reward of a pool with total stake $\sigma$ and allocated pool leader stake $\lambda$. It holds $\sum_j r(\sigma_j, a_{j,j}) \le R$.
- Potential profit of a saturated pool with allocated pool leader stake $\lambda$ and cost $c$, $P(\lambda, c) = r(\beta, \lambda) - c$.
- We order the players according to $P(s_i, c_i)$. Player $i$ is the player with the $i_{th}$ highest $P(s_i, c_i)$.
- $\vec{S}^{(\vec{m},\vec{\lambda})} = (\vec{a}_i)_{i=1}^{n}$, joint strategy regarding allocation given $(\vec{m}, \vec{\lambda})$. $a_{i,i} \in \{0, \lambda_i\}$ and $S_i^{(\vec{m},\vec{\lambda})} = \vec{a}_i$.
- $\alpha \in [0, \infty)$: parameter that can be adapted to trade between efficiency and Sybil resilience. Note that the total rewards $R$ and $\alpha$ should be selected such as it holds also $P(s_{k+1}, c_{k+1}) > 0$.
- $u_{j,i}(\vec{S}^{(\vec{m},\vec{\lambda})})$, (myopic) utility that player $j$ gets from pool $\pi_i$.

$$u_{j,i} = \begin{cases} 0 & r(\sigma_i, a_{i,i}) \le c_i \\ \frac{a_{j,i}}{\sigma_i} \cdot (r(\sigma_i, a_{i,i}) - c_i) \cdot (1 - m_i) & \text{otherwise} \end{cases}$$

- $u_{i,i}(\vec{S}^{(\vec{m},\vec{\lambda})})$, (myopic) utility that player $i$ gets from their own pool $\pi_i$.

$$u_{i,i} = \begin{cases} r(\sigma_i, a_{i,i}) - c_i & r(\sigma_i, a_{i,i}) \le c_i \\ (m_i + (1 - m_i) \cdot \frac{a_{i,i}}{\sigma_i}) \cdot (r(\sigma_i, a_{i,i}) - c_i) & \text{otherwise} \end{cases}$$

- $u_j(\vec{S}^{(\vec{m},\vec{\lambda})})$, total (myopic) utility of player $j$: $u_j(\vec{S}^{(\vec{m},\vec{\lambda})}) = \sum_{i=1}^{n} u_{j,i}(\vec{S}^{(\vec{m},\vec{\lambda})})$.
- Non-myopic utility is defined in the same way as myopic utility but by using non-myopic stake $\sigma^{\text{NM}}$ instead. Refer to discussion above and Definitions 7, 8.
- $(x)^+ = \max\{0, x\}$

Figure 1: Notations and concepts introduced.

the reward $R$ is distributed to the pools and pool members, and *the central issue of this work is to determine reward schemes with desired properties.*

We assume that the stakeholders are rational in the sense that they want to maximize their utility and that there are *no externalities*, i.e., outside factors that affect the reward of the pool and the players.

Our primary objective is to incentivize the stakeholders to form a certain number of pools (smaller than the number of players). We further want no pool to have a disproportionally large size, so that no group can exert disproportionally large influence. Ideally, we want to find a reward scheme that, at equilibrium, leads to the creation of many almost equal-stake pools independently of (i) number of players (ii) the distribution of stake and costs (iii) the degree of concurrency in selecting a strategy. This seems like an impossible task[5], so we have to settle for solutions that achieve the above goals approximately under some natural assumptions about the distribution of stake and costs and about the equilibria selection dynamics.

We summarize the model here. Formal definitions of the concepts follow next.

**Reward sharing schemes (RSS) for stake pools.** The class of reward sharing schemes we investigate is parameterised by a function $r : [0,1]^2 \to \mathbb{R}_{\geq 0}$ and operates as follows.

- The reward scheme distributes a total fixed amount $R$ to the pools according to their stake $\sigma_i$ and the stake of their pool leader $a_{i,i}$. In particular pool $\pi_i$ gets reward $r(\sigma_i, a_{i,i})$ with $\sum_i r(\sigma_i, a_{i,i}) \leq R$. Note that we don't have to distribute the whole amount $R$. Formally, the function $r(\cdot, \cdot)$ takes the stake of a pool and the stake of the pool leader allocated to this pool and returns the payment for this pool so that: $\sum_i r(\sigma_i, a_{i,i}) \leq R$.

- $r(0,0) = 0$, which means that a pool with no stake will get zero rewards.

- The reward $r(\sigma_i, a_{i,i})$ of each pool $\pi_i$ is shared among its pool leader and its stakeholders. This may be done in a number of ways but in any case, the pool leader should get an amount $c_i^- = \min(c_i, r(\sigma_i, a_{i,i}))$ to cover the declared cost for running the pool. We will focus our investigation on reward schemes that are *proportional*, i.e., those schemes that have the property that the ratio of the rewards obtained by stakeholder $j_1$ over the rewards of stakeholder $j_2$ in pool $\pi_i$ equals $a_{j_1,i}/a_{j_2,i}$, with the only exception being for pool leaders who may be considered for additional rewards.

**The stake pools game and utility function.** Based on a reward scheme as described above, we can define the stake pools game where the strategies of the players are their allocations of their stake to their own as well as the other available pools. In this game each player $i$ tries to maximize his utility. The rewards of a pool $\pi_i$ are $r(\sigma_i, a_{i,i})$ and the cost the pool leader/operator incurs for running this pool is $c_i$. The pool operator gets his cost reimbursed, apart from that, all rewards are split proportional to stake. So if a player i with cost $c_i$ runs a pool with total stake $\sigma_i$, his utility $u_{i,i}$ from this pool $\pi_i$ is

$$u_{i,i} = \begin{cases} r(\sigma_i, a_{i,i}) - c_i & \text{for } r(\sigma_i, a_{i,i}) \leq c_i, \\ \frac{a_{i,i}}{\sigma_i} \cdot (r(\sigma_i, a_{i,i}) - c_i) & \text{otherwise,} \end{cases}$$

and a player $j \neq i$ delegating stake $a_{j,i}$ to that pool $\pi_i$ will get rewards

$$u_{j,i} = \begin{cases} 0 & \text{for } r(\sigma_i, a_{i,i}) \leq c_i, \\ \frac{a_{j,i}}{\sigma_i} \cdot (r(\sigma_i, a_{i,i}) - c_i) & \text{otherwise} \end{cases}$$

from that pool. We define the utility of each player $j$ to be $u_j = \sum_i u_{j,i}$. Given the above, the hard question is to define the reward sharing scheme, and importantly $r(\cdot, \cdot)$, so that the underlying stake pools game has Nash equilibria that meet (at least) our primary objective: having a large number of active pools.

---

[5] Actually, here is a simple reward scheme that achieves all goals: give no reward to the pools, unless there are many equal-stake pools, in which case each pool gets reward $R/k$. However, we are interested in reward schemes that can lead to a good Nash equilibrium starting at the state in which all players play no-participation and following natural symmetric, almost myopic dynamics, such as repeatedly having a random player playing best-response.

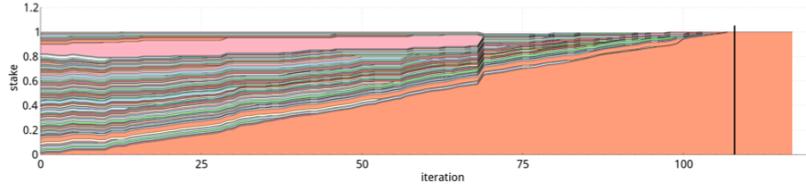

Figure 2: Example dynamics for the fair reward sharing scheme ($c \in [0.001, 0.002]$) showing centralisation after about 100 iterations with $n = 100$ players. Initially, the players are "maximally decentralzed". Here and in all following similar diagrams, the vertical line indicates the time when equilibrium is reached.

## 2.2 Fair RSS's and their Failure to Decentralise

In this subsection we will show that if we use a "fair" reward sharing scheme, then we will end up in an equilibrium with at most one pool, which means that this scheme fails our decentralization objective.

Specifically consider the fair allocation that sets $r(\sigma_i, a_{i,i}) = \sigma_i \cdot R$, i.e., pools are rewarded proportionally to their size. For simplicity we will take $R = 1$. (Note that if we consider $R = 1$ then all the costs are between zero and one.) Moreover, we will assume that all pool participants are also treated fairly receiving rewards proportionally to the stake they have delegated in the pool of their choice.

We prove the following (see Appendix A.1 for the proof) the following theorem:

**Theorem 2.** *Given the above reward sharing scheme: (I) There is no equilibrium where more than one pool is created.*

*(II) If there exists i such that $s_i > c_i$ then the only equilibria are the following: there exists just one pool, say $\pi_i$ and it holds (i) $c_i \leq 1$ and (ii) $s_j \cdot c_i \leq c_j$ for each member j of this pool (iii) all players have delegated their stake to $\pi_i$.*

**Experimental results – dynamics**

Given the above theorem, we then experimentally investigate *how fast such systems centralize*. We use three different initial states for these experiments:

1. "Maximally decentralized", where every player whose cost $c_i$ is lower than his stake $s_i$ runs a pool and all other players are passive.

2. "Inactive", where no player runs a pool.

3. "Nicely decentralized", where ten players run a pool, and the others delegate to these pools in a way that makes them all equally big.

Our experiments show that the convergence to the results predicted by the theory is fast: If at least one player has stake greater than cost and hence runs a pool, all players will end up delegating all their stake to this single pool ending up in a "dictatorial" single pool configuration. The simulation in the experiment has players selected at random taking turns and playing best-response attempting to maximise their utility. More details regarding how the experiments are executed refer to Section 7 where we overview our experiments.

In Figures 2, 3 and 4 we present a graphical representation of the experiments. Different colors correspond to different pools. The $x$-axis represents time while the $y$-axis the stakeholders. Costs are uniformly selected in the specified range. Stake is following a Pareto distribution.

In the following theorem (i) we generalise the impossibility result to the case of any function $r$ for which $(r(\sigma, \lambda) - c)/\sigma$ is strictly increasing in $\sigma$ and (ii) we prove that there are configurations for which there is no equilibrium with a number of pools smaller than the number of players in the case of a strictly decreasing $(r(\sigma, \lambda) - c)/\sigma$ in $\sigma$.

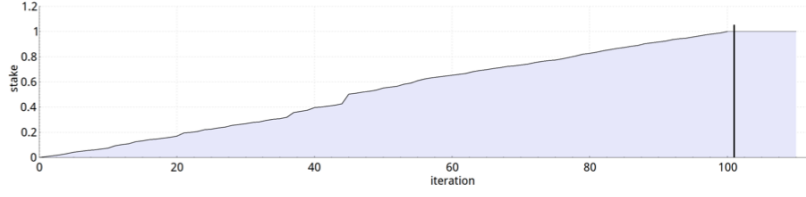

Figure 3: Example dynamics for the fair reward sharing scheme ($c \in [0.001, 0.002]$) showing centralisation after about 100 iterations with $n = 100$ players. Initially, no stake-pools exist.

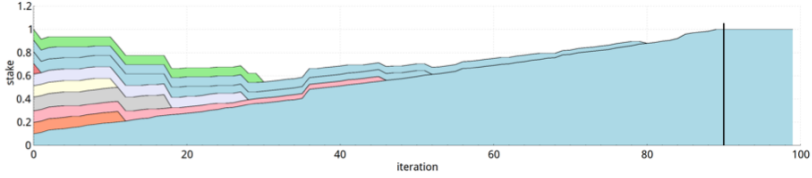

Figure 4: Example dynamics for the fair reward sharing scheme ($c \in [0.001, 0.002]$) showing centralisation after about 100 iterations with $n = 100$ players. Initially, the players are "nicely decentralized".

**Theorem 3.** *I) If $\dfrac{r(\sigma, \lambda) - c}{\sigma}$ as a function of $\sigma$ is strictly increasing in $\sigma \in (0, 1]$ then there is no equilibrium with more than one pool. Note that a fair reward function $r(\sigma, \lambda) = \sigma$ is such an example.*

*II) If $r(\sigma, \lambda) = r(\sigma)$ a continuous and strictly increasing function on $\sigma$ and $\dfrac{r(\sigma, \lambda) - c}{\sigma}$ as a function of $\sigma$ is strictly decreasing in $\sigma \in [\sigma_0, 1]$, where $\sigma_0$ such that $r(\sigma_0) - c > 0$, then there is an assignment of costs and stakes to the players such that there is no equilibrium with fewer than n pools where n the number of players. We will assume for the proof that each player can delegate to a pool stake at least $\dfrac{s_{min}}{f}$ where $f \in (1, \infty)$ and $s_{min}$ the minimum stake among all the players.*

For the proofs see in Appendix A.2.

## 2.3 RSS with Cap and Margin

Motivated by the failure of the fair reward sharing scheme, in this section we will put forth a wider class of reward sharing schemes that fare better (as we will demonstrate) in terms of incentivizing players to create many pools of similar size.

Our first key observation for a reward function to have better potential for decentralization is that while it should be increasing for small values of the pool's stake, something that will incentivize players to join together in pools to share their costs, the rewards should plateaux after a certain point in order to discourage the creation of large pools, or equivalently to incentivize the breakup of large pools into smaller ones. This suggests that rewards will be *capped*.

Our second observation is that it is sensible to treat pool leaders in a preferential way with respect to rewards. Recall that in the case when the rewards of the pool are more than the cost, the cost is subtracted from the rewards of the pool and, if we treat everyone proportionally, the pool leader should get the same rewards as a pool member having delegated the same stake to the pool. On the other hand, in the case when the pool does not get enough rewards to compensate its operational cost then the difference is paid by the pool leader. So the pool leader bears an extra risk compared to regular pool members and it makes sense to be compensated for that. Thus, in our reward scheme we will consider that the pool leader can ask for an extra reward compared to the other members. This reward will be a fraction of the pool's profit and this fraction will be denoted by the *margin* value $m$. The margin will be part of the strategy of potential pool leaders.

**Reward sharing scheme with cap and margin**. A reward scheme for stake pools that incorporates the above features will be called *reward sharing scheme with cap and margin*. Formally :

**Definition 2** (Reward sharing schemes with cap and margin)**.** A reward sharing scheme with cap and margin is a reward sharing scheme that (1) is parameterised by a function $r : [0,1]^2 \to \mathbb{R}_{\geq 0}$ (that takes as input the stake $\sigma_i$ of a pool $\pi_i$ and the stake $a_{i,i}$ of the pool leader allocated to this pool and returns the total reward for this pool) and a value $k \in \mathbb{N}$ and satisfies the following properties:

- (as before) $\sum_{i=1}^{n} r(\sigma_i, a_{i,i}) \leq R$, where $R$ the total rewards.

- (as before) $r(0,0) = 0$.

- $\frac{d[(r(\sigma,\lambda)-c)\cdot \frac{1}{\sigma}]}{d\sigma} > 0$, when $\sigma \leq \beta \stackrel{\text{def}}{=} \frac{1}{k}$. This means that the reward function is increasing for small values of pool's stake to incentivize players to join together in pools to share the cost.

- $\forall \lambda\ r(\sigma, \lambda) = r(\beta, \lambda)$ when $\sigma > \beta$. This means that the reward function is constant for large values of the pool's stake to discourage the creation of large pools.

(2) the reward $r(\sigma_i, a_{i,i})$ of each pool $\pi_i$ is shared among its pool leader and its stakeholders. The pool leader gets an amount $c_i^- = \min(c_i, r(\sigma_i, a_{i,i}))$ to cover the declared cost for running the pool. A fraction $m_i$ of the remaining amount $(r(\sigma_i, a_{i,i}) - c_i^-)$ is the pool leader compensation for running the pool. This fraction is referred to as *margin*. The rest $(1 - m_i) \cdot (r(\sigma_i, a_{i,i}) - c_i^-)$ is distributed to the stakeholders of the pool, including the pool leader, *proportionally to their contributed stake*.

To analyze the outcome of a reward scheme, we need to define the game induced by it, which in turn depends on our assumptions about how far-sighted the players are when calculating their best response. We analyze the natural assumption that each player computes their utility using the estimated final size of the pools (under the assumption that the other players act in the same way). The utility of the players in this setting depends on the *desirability* $D_j(\vec{S}^{(\vec{m},\vec{\lambda})}) = (1-m_j)P(\lambda_j, c_j)^+$ of pool $\pi_j$, where $P(\lambda_j, c) = r(\beta, \lambda_j) - c$ is the potential profit of the pool when it is saturated. Each player ranks the pools according to their desirability and computes the expected stake $\sigma_j^{\text{NM}}$ of them (this is related to the *non-myopic stake*, see definition 7), which is either $\max(\beta, \sigma_j)$, when the pool is ranked among the $k$ most desirable pools, or simply $\lambda_j + a_{i,j}$, when the pool is not very desirable and the player expects to be alone with the pool leader. With this, we see that the non-myopic utility that the player gets by committing stake $a_{i,j}$ to a saturated pool $\pi_j$ is $D_j(\vec{S}^{(\vec{m},\vec{\lambda})})a_{i,j}/\sigma_j^{\text{NM}}$. The utility of the pool leaders is computed accordingly.

## 3 Formal Treatment of the Stake Pools Game

**The stake pools game with cap and margin.** Without loss of generality we assume that every player can be the leader of only one pool and each player has stake at most $\beta = 1/k$; players with stake more than $\beta$ or wishing to create more than one pool can be thought of as a strategic coalition of players which we analyse in Section 4 where we consider Sybil attacks of this nature. Below, we will use the notation: $(x)^+ = \max(0, x)$, and $[n] = \{1, \ldots, n\}$.

**Definition 3** (Strategy of a player)**.** The strategy of a player $i$ has two parts:

- $(m_i, \lambda_i)$, where $m_i \in [0,1]$ is the margin and $\lambda_i$ the stake that player $i$ will commit if he activates his own pool.

- $S_i^{(\vec{m},\vec{\lambda})} = \vec{a}_i^{(\vec{m},\vec{\lambda})}$ that is the allocation of player $i$' stake given $(\vec{m}, \vec{\lambda})$. When the $(\vec{m}, \vec{\lambda})$ can be inferred from the context we will use $\vec{a}_i$ for simplicity. $a_{i,j} \in [0,1]$ denotes the stake that player $i$ allocates to pool $\pi_j$ so that his total allocated stake is $\sum_{j=1}^{n} a_{i,j} \leq s_i$. This allows for stake $s_i - \sum_{j=1}^{n} a_{i,j}$ of the player to remain unallocated. In addition $a_{i,i}^{(\vec{m},\vec{\lambda})} \in \{0, \lambda_i\}$.

**Definition 4** (Pools)**.** Given a joint strategy $\vec{S}^{(\vec{m},\vec{\lambda})}$, the stake allocated to a pool $\pi_j$ is denoted by $\sigma_j(\vec{S}^{(\vec{m},\vec{\lambda})})$, or simply $\sigma_j$ for a less cluttered notation. A pool $\pi_j$ is called *active* when player $j$ allocates non-zero stake to it, that is, $a_{j,j} = \lambda_j > 0$. Note that only player $j$ can activate pool $\pi_j$. If a pool $\pi_j$ is active its stake is $\sigma_j = \sum_{i=1}^n a_{i,j}$, otherwise we assume that $\sigma_j = 0$. A pool is called *saturated* when its stake is at least $\beta$.

The restriction that only player $j$ can activate pool $\pi_j$, by allocating non-zero stake to it, is necessary to prevent other players to force player $j$ pay the cost $c_j$ of operating the pool without consenting to open the pool.

**Non-myopic utility for reward sharing schemes with cap and margin.** Recall that the strategy of player $i$ is either to become a pool leader with margin $m_i$ by committing stake $\lambda_i$ and/or to delegate his stake to other pools.

A crucial observation is that if we extend directly the utility we have defined in the game for stake pools so that it includes margin, then in the game defined by the above set of strategies, the notion of Nash equilibrium does not match the intuitive notion of stability that an equilibrium is supposed to provide. Note that, in the context of a Nash equilibrium, when players try to maximize utility, they play in a myopic way, which means that they decide based on the current size of the pools and they do not take into account what effect their moves have on the moves of the other players and thus, ultimately, in the eventual size of the pools. To see the issue, suppose that we have reached a Nash equilibrium in this game, that is, a set of strategies from which no player has an incentive to deviate unilaterally. The obvious problem is that at Nash equilibrium *all margins will be* 1. This is so, because by the definition of the Nash equilibrium the other players will keep their current strategy, and the best response of a pool leader is to select the maximum possible margin. Thus, if there is room to increase the margin, the strategy cannot be a Nash equilibrium and hence the only equilibrium, if it exists, will exhibit all margins to be to their maximum value 1. There are two problems here: first we definitely don't want the margins to be 1, and second, such an outcome is not expected to be a stable solution anyway! (In a sense contradicting the intuitive notion of what a Nash equilibrium is supposed to offer). If all margins are 1, a non-myopic player (a forward-looking player who tries to predict the final size of the pools after the other players play) who is not a pool leader can start a new pool with smaller margin which will attract enough stake to make it profitable.

For these reasons, in order to analyse our *reward sharing schemes with cap and margin* we will use a natural *non-myopic* type of utility which enables the players to be more far-sighted. Thus, in the analysis, players will not consider myopic best responses but non-myopic best responses. Specifically, *a player computes his utility using the estimated final size of the pools instead of the current size of the pools*. The estimated final size is either the stake that the pool leader has allocated to this pool or the size of a saturated pool. The latter is used when the pool is currently ranked to belong among the most desirable pools and the former when the pool does not belong among them. It follows that a non-myopic player that considers where to allocate his stake, would want to rank the pools with respect to the estimated reward at the Nash equilibrium. But this reward is not well-defined because the Nash equilibrium depends on the decisions of the other players. It makes sense then to use a crude ranking of the pools. Such a ranking can be based on the following thinking: "An unsaturated pool" where I will place my stake will also be preferred by other like-minded players if it has relatively low margin and cost, and substantial stake committed by the pool leader (the last one is essential only when $\alpha \neq 0$) so the pool will become saturated. So, I will assume that the stake of the pool is actually $\beta$. On the other hand, if a pool has relatively high margin and cost and/or not substantial stake committed by the pool leader will not grow and will lose also its members as other unsaturated pools offer better combination of margin and cost. This motivates the following ranking of pools:

**Definition 5** (Desirability and Potential Profit)**.** The potential profit of a saturated pool with allocated pool leader stake $\lambda$ and cost $c$ is $P(\lambda, c) = r(\beta, \lambda) - c$. Given a joint strategy $\vec{S}^{(\vec{m},\vec{\lambda})}$, we define the desirability of a pool $\pi_j$

$$D_j(\vec{S}^{(\vec{m},\vec{\lambda})}) = \begin{cases} (1-m_j)P(\lambda_j, c_j) & \text{if } P(\lambda_j, c_j) \geq 0 \\ 0 & \text{elsewhere} \end{cases} \quad (3)$$

Note that the desirability of a pool depends on its margin, the stake of the pool leader allocated to this pool and its cost.

**Definition 6** (Ranking). Given a joint strategy $\vec{S}^{(\vec{m},\vec{\lambda})}$, the rank of a pool $\pi_j$ denoted by $\text{rank}_j(\vec{S}^{(\vec{m},\vec{\lambda})})$ is its ranking with respect to the desirability $D_j(\vec{S}^{(\vec{m},\vec{\lambda})})$. The maximum desirability gets rank 1, the second maximum desirability gets rank 2, etc. Again to get a less cluttered notation, we will write $\text{rank}_j$ instead of $\text{rank}_j(\vec{S}^{(\vec{m},\vec{\lambda})})$ whenever the joint strategy $\vec{S}^{(\vec{m},\vec{\lambda})}$ can be inferred from the context. Ties break according to the potential profit, specifically the pool with the higher potential profit will be ranked higher; (with higher we mean smaller rank) for convenience we assume that all potential profit values are distinct. The $k$ most desirable pools will be these ones with rank smaller or equal to $k$.

Given the ranking, we define the non-myopic stake of a pool to be either the stake allocated by the pool leader or the size of a saturated pool. The first one is used when the pool does not belong to the $k$ most desirable pools and the second one when the pool is among them.

**Definition 7** (Non-myopic stake). The non-myopic stake of pool $\pi_j$ is defined as

$$\sigma_j^{\text{NM}}(\vec{S}^{(\vec{m},\vec{\lambda})}) = \begin{cases} \max(\beta, \sigma_j) & \text{if } \text{rank}_j \leq k \\ a_{j,j} & \text{otherwise.} \end{cases} \quad (4)$$

To simplify the notation we use $\sigma_j^{\text{NM}}$ instead of $\sigma_j^{\text{NM}}(\vec{S}^{(\vec{m},\vec{\lambda})})$, $\sigma_j$ instead of $\sigma_j(\vec{S}^{(\vec{m},\vec{\lambda})})$, $\text{rank}_j$ instead of $\text{rank}_j(\vec{S}^{(\vec{m},\vec{\lambda})})$ and $a_{j,j}$ instead of $a_{j,j}(\vec{S}^{(\vec{m},\vec{\lambda})})$.

**Definition 8** (Non myopic utility). The utility $u_i(\vec{S}^{(\vec{m},\vec{\lambda})})$ of player $i$ from being a member of pool $\pi_j$ with non myopic stake $\sigma_j^{\text{NM}}$ is

$u_{i,j}(\vec{S}^{(\vec{m},\vec{\lambda})}) =$

$$\begin{cases} 0 & \pi_j \text{ is inactive } (a_{j,j} = 0) \\ (1-m_j)\left(r(\beta, \lambda_j) - c_j\right)^+ \frac{a_{i,j}}{\sigma_j^{\text{NM}}} & \text{rank}_j \leq k \wedge a_{j,j} \neq 0 \\ (1-m_j)\left(r(\lambda_j + a_{i,j}, \lambda_j) - c_j\right)^+ \frac{a_{i,j}}{\lambda_j + a_{i,j}} & \text{otherwise.} \end{cases}$$

The utility $u_j(\vec{S}^{(\vec{m},\vec{\lambda})})$ that the pool leader $j$ gets from pool $\pi_j$ is

$u_{j,j}(\vec{S}^{(\vec{m},\vec{\lambda})}) =$

$$\begin{cases} 0 & \pi_j \text{ is inactive } (a_{j,j} = 0) \\ r(\sigma_j^{\text{NM}}, \lambda_j) - c_j & r(\sigma_j^{\text{NM}}, \lambda_j) - c_j < 0 \wedge a_{j,j} \neq 0 \\ (r(\sigma_j^{\text{NM}}, \lambda_j) - c_j)\left(m_j + (1-m_j)\frac{\lambda_j}{\sigma_j^{\text{NM}}}\right) & \text{otherwise.} \end{cases}$$

The utility of player $i$ is the sum of the utilities coming from all pools in which he participates as a pool leader or a pool member: $u_i(\vec{S}^{(\vec{m},\vec{\lambda})}) = \sum_{j=1}^n u_{i,j}(\vec{S}^{(\vec{m},\vec{\lambda})})$.

# 4 A Sybil Resilient Reward Sharing Scheme

In this section, we first outline the motivation behind our choice of the parameterized reward function.
**Motivating our solution.** We propose a reward sharing scheme with cap and margin cf. Definition 2. To motivate this choice, let us first consider a reward function $r(\sigma, \lambda) = r(\sigma, 0)$ that depends only on the total stake $\sigma$ of the pool (note we assume without loss of generality that the stake of any agent or pool belongs to $(0, 1)$ and represents the fraction of the total stake controlled by the specific entity) and it is independent of the stake $\lambda$ of the pool leader. The natural choice is to select $r(\sigma, 0)$ proportional to $\sigma$, which has the nice property that it rewards all players proportionally to their stake. However as we have seen already in Section 2.2, it leads to dictatorial equilibria in which a single pool is created. (Note that the cost of running a stake pool remains the same regardless its size). Moreover, it is clear that if we want to achieve a target number of pools, say $k$, it is clear any similar reward scheme cannot achieve this goal since it is independent of the target $k$. This motivates a simple modification of this reward scheme which goes a long way in meeting this target. Consider the modification

$$r(\sigma, 0) \sim \min\{\sigma, \beta\},$$

where $\beta$ is a constant (this is the *cap*) and $\sim$ indicates proportionality with a multiplier that guarantees that the total reward is sufficient to pay all pools[6] (see Figure 5).

Recall a pool is *saturated* when its total stake $\sigma$ is at least $\beta$, so we can say that such a capped reward function discourages oversaturated pools. By setting $\beta = 1/k$, this reward scheme seems to provide the right incentives to create pools of size up to $\beta = 1/k$, which naturally leads to $k$ pools of equal size. However, this picture is to a large extent misleading because the usual myopic best-response dynamics creates a single pool instead of $k$, because even with this reward function, for a pool member, a saturated pool is preferable to a pool whose reward is mainly used to cover the cost of its leader. The good news is that, as we will show, dynamics of non-myopic best response achieves the goal by leading to an equilibrium of $k$ pools of equal size, given a reasonable definition of an appropriate non-myopic notion of utility.

To evaluate the quality of a reward scheme, we should compare the resulting equilibrium with an optimal solution. An optimal solution when all participants act honestly and selfishly is to have $k$ pools of equal size that are run by pool leaders of minimal cost. This would make the system efficient, in both computational and economic sense. But besides efficiency, we want the system to withstand attacks from some players that try to run many pools, even at a loss.

**Sybil behavior and resilience.** In particular we want to disincentivize Sybil strategies [14]) that create multiple identities declaring potentially lower costs for each one. We distinguish two types of Sybil behaviors: the first one captures a non-utility maximizer who wants to control 50% of the system. Such level of control enables a party to perform double spending attacks on the blockchain or arbitrarily censor transactions. The second type of Sybil behavior is that of a utility maximizer that creates multiple identities with their corresponding stake-pools sharing the same server back-end and thus also the operational costs. Such a player limits decentralisation by reducing the number of independent server deployments that provide the service. Observe that this also can include coalitions of players that decide to act as one. Such behavior cannot be excluded in the anonymous setting that we operate. The best possible that we can hope for is to lower bound the stake of the Sybil player to be *linear* in the number of identities that it creates. We analyse the Sybil resilience of a reward sharing scheme by estimating the minimum stake $s_{\min}$ needed for the Sybil behavior to be effective.

To address this issue we design a reward sharing scheme that guarantees that players can attract stake from other players only if they commit substantial stake to their own pool. This is precisely the reason for considering reward functions that depend, besides the total stake of the pool, on the stake of the pool leader.

---

[6]A smooth function that approximates this reward function may be preferable to improve the dynamics of convergence to equilibrium.

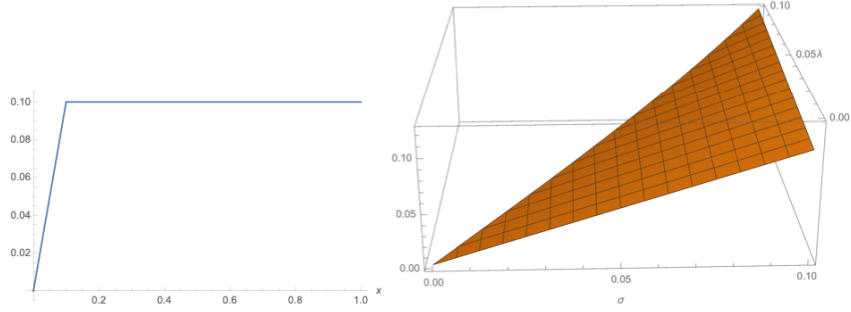

Figure 5: Reward function for $\beta = 1/10$ with $\alpha = 0$ (top) and $\alpha = 1/4$ (bottom).

Ideally, we want the pools to be created by the players ranked best according to $\alpha s_j - c_j$ (a linear combination of their stake $s_j$ and their cost $c_j$), where $\alpha$ is a nonnegative parameter that can be adapted to trade between efficiency and Sybil resilience. By selecting $\alpha = 0$ we get the most efficient solution, and on the other extreme, by selecting a very large $\alpha$, we can obtain a potentially inefficient solution in which the pool leaders might be the $k$ "wealthiest" but the Sybil resilience of the system improves.

The objective is to design a reward scheme that provides incentives to obtain an equilibrium that compares well with the above optimal solution. On the other hand, we feel that it is important that the mechanism is not unnecessarily restrictive and all players have the "right" to become pool leaders.

The natural way to accomodate this in our scheme, would be to use the above reward function but apply it to $\sigma + \alpha \lambda$, a weighted sum of the total pool stake $\sigma$ and the allocated pool leader stake $\lambda$. With this in mind, the reward function becomes $r(\sigma, \lambda) \sim \min\{\sigma, \beta\} + \alpha \lambda$. Again this reward function goes some way towards meeting the objective but the best response dynamics, even non-myopic best response dynamics, do not lead to equilibria that resemble the optimal solution and in particular, it may create pools of very large size. The reason is that the influence of the stake $\lambda$ of the pool leader when a pool is still small is very significant. Given that the ideal size of the pool is $\beta$, one way to alleviate this effect is to change the influence factor $\alpha$ to be proportional to the stake that the pool has already attracted, that is to change the influence factor to $\alpha' = \alpha \frac{\sigma - \lambda}{\beta}$. This creates the (more minor) problem that the influence factor will not be the same for all pools, which is quite desirable when a parameterisation is attempted and the value of $\alpha$ will be used to control Sybil attacks. The final touch in the reward function which resolves this issue is to make the influence of the stake of the pool leader on the factor $\alpha'$ to disappear when the pool has the desired size of $\beta$. The resulting reward function described briefly in the informal theorem of the introduction (Definition 1) is defined and analyzed in the rest of the current section.

### 4.1 Our RSS construction

Given our target number of pools $k$, we define the reward function $r_k : [0,1]^2 \to \mathbb{R}_{\geq 0}$ of a pool $\pi$ with stake $\sigma$ and pool leader's allocated stake $\lambda$ as follows:

$$r_k(\sigma, \lambda) = \frac{R}{1+\alpha} \cdot [\sigma' + \lambda' \cdot \alpha \cdot \frac{\sigma' - \lambda' \cdot (1 - \sigma'/\beta)}{\beta}],$$

where $\lambda' = \min\{\lambda, \beta\}$, $\sigma' = \min\{\sigma, \beta\}$ and $\beta, \alpha$ are fixed parameters. A natural choice is $\beta = 1/k$, where $k$ is the target of number of pools. For simplicity we will write $r$ instead of $r_k$.

We have: $\alpha \in [0, \infty)$, $k \in \mathbb{N}, (k < n)$ and $R \in \mathbb{R}$. Note that the total rewards $R$ and $\alpha$ should be selected such as it holds also $P(s_{k+1}, c_{k+1}) > 0$.

The next proposition shows that the proposed function is suitable for a *reward sharing scheme with cap and margin*.

**Proposition 1.** *The function $r(\cdot,\cdot)$ satisfies the properties of a reward sharing scheme with cap and margin, cf. Definition 2.*

*Proof.* It holds $\forall i \; r(\sigma_i, a_{i,i}) \geq 0$, as $a'_{i,i} \leq \sigma'_i$ and also:

1. $\sum_{i=1}^n r(\sigma_i, a_{i,i}) \leq R$, as $\frac{\sigma'_i - a'_{i,i} \cdot \frac{(\beta - \sigma')}{\beta}}{\beta} \leq 1$ and $\sum_{i=1}^n [\sigma_i + a_{i,i} \cdot \alpha] = \sum_{i=1}^n \sigma_i + \alpha \cdot \sum_{i=1}^n a_{i,i} \leq 1 + \alpha$.

2. $r(0,0) = 0$.

3. When $\sigma \leq \beta$ it holds: $\frac{d[r(\sigma,\lambda) - c) \cdot \frac{1}{\sigma}]}{d\sigma} > 0$.

4. $\forall \lambda \; r(\sigma, \lambda) = r(\beta, \lambda)$, when $\sigma > \beta$ because we have $\sigma' = \min\{\sigma, \beta\}$.

This completes the proof. □

## 4.2 Perfect Strategies

We define a class of strategies and we prove that they are *Nash equilibria* of our game (Theorem 4). This class has the following characteristics: exactly $k$ pools of equal size are created and the pool leaders are the players with the highest value $P(s,c)$ (when $\alpha = 0$ those are the players with the smallest cost). Recall that the players are ordered in terms of potential profit, e.g., player 1 is the player with the highest $P(s_i, c_i)$. Recall also that players decide to create or not a pool and how much stake they will allocate to other pools. In addition they decide a margin for their potential pool.

**Definition 9** (Perfect strategies)**.** We define a class of strategies, which we will call perfect. The margins are

$$m_j^* = \begin{cases} 1 - \frac{P(s_{k+1}, c_{k+1})}{P(s_j, c_j)} & \text{when } j \leq k \\ 0 & \text{otherwise,} \end{cases}$$

the stake allocated by each pool leader to their own pool is equal to their whole stake and the allocations are such that each of the first $k$ pools has stake $\beta$.

Note that when $j \leq k$ it holds $\text{rank}_j \leq k$.

The following proposition gives the utilities at perfect strategies and it follows directly from Definition 8 of the non-myopic utilities of pool members and pool leaders and our reward function described in this section.

**Proposition 2.** *In every perfect strategy, (i) the utilities of the players are:*

$$u_i = P(s_{k+1}, c_{k+1})\frac{s_i}{\beta} + (P(s_i, c_i) - P(s_{k+1}, c_{k+1}))^+, \tag{5}$$

*and (ii) the desirability of the first $k + 1$ players is the* same *and equal to $P(s_{k+1}, c_{k+1})$.*

To justify the proposition note that all the players get a fair reward, in the sense that it is a constant $P(s_{k+1}, c_{k+1})/\beta$ times their stake, with the exception of each pool leader $i$, who gets an additional reward $P(s_i, c_i) - P(s_{k+1}, c_{k+1})$. This additional reward can be viewed as a bonus for the efficiency and security that the pool leader brings to the system. We will show that every perfect strategy is a Nash equilibrium of the game with the defined utilities.

**Theorem 4.** *Every perfect strategy is a Nash equilibrium.*

Before presenting the proof of the theorem we start with some definitions and preliminary results.

**Definition 10** (Desirability of a player)**.** Desirability of a player will be the desirability of their pool. If they do not have one, their desirability will be the desirability of a hypothetical pool with their cost, the margin they have chosen and their personal stake.

Note that for uniformity we assume that all the players decide a margin even if they do not create a pool. In addition, when we rank the pools in this subsection, we will take into account also the hypothetical pools described above. Ties break in favor of potential profit. In the two-stage game that we examine in Section 5 we remove these assumptions (regarding hypothetical pools and ties as (i) we do not take into account non active pools in the ranking because we consider their desirability as zero (ii) ties in ranking break arbitrarily).

The following lemma is very useful and its proof follows directly from the definition of the reward function.

**Lemma 1.** *The quantity $(r(x, s_j) - c_j)/x$ as a function of $x$ is increasing in $(0, \beta)$ and, if it is positive, decreasing in $(\beta, \infty)$. Its maximum is achieved at $x = \beta$.*

The following lemma gives an upper bound on the utility of pool members. We will give an equilibrium that matches this upper bound.

**Lemma 2.** *In every joint strategy in which some player $j$ is not a pool leader, their utility is at most $\max_l D_l \cdot (s_j/\beta)$, where $\max_l D_l$ is the maximum desirability among all players.*

*Proof.* It suffices to show that player $j$ gets at most $D_l \frac{a_{j,l}}{\beta}$ from every pool $l$. The lemma follows directly from this by summing for all $l$: $\sum_l D_l \frac{a_{j,l}}{\beta} \leq \max_l D_l \sum_l \frac{a_{j,l}}{\beta} = \max_l D_l \frac{s_j}{\beta}$. The argument that for every pool $l$, player $j$ gets at most $D_l \frac{a_{j,l}}{\beta}$ follows directly from the definition of the utility of pool members when we consider the two cases depending on whether $\text{rank}_l$ is at most $k$ and more than $k$.

Specifically, when $\text{rank}_l \leq k$, by the definition of the utility of pool members, the utility to player $j$ from pool $l$ is $D_l a_{j,l} / \sigma_l^{NM} \leq D_l a_{j,l} / \beta$.

When $\text{rank}_l > k$, his utility is given by

$$(1 - m_l)\left(r(\lambda_l + a_{j,l}, \lambda_l) - c_l\right)^+ \frac{a_{j,l}}{\lambda_l + a_{j,l}}$$
$$\leq (1 - m_l)\left(r(\beta, \lambda_l) - c_l\right)^+ \frac{a_{j,l}}{\beta}$$
$$= D_l \frac{a_{j,l}}{\beta},$$

where the inequality comes from Lemma 1. □

We are now ready to present the proof of the Theorem.

*Proof.* (of Theorem 4) We first consider the simplified setting where players are mutually exclusively pool leaders or pool members.

Consider first a player $j$ with rank at most $k$. This player is a pool leader of a pool of size $\beta$. We show that none of the possible responses improves their utility:

- Suppose that the player decreases their margin. This increases their desirability so that the new rank is still one of the first $k$ ranks. Since the non-myopic stake remains the same, this move will decrease the utility of the player.

- Suppose that the player increases their margin. Since before the change the first $k + 1$ players have the same desirability, the player's desirability drops and the rank becomes larger than $k$. As a result the player will be alone in a pool and their utility can only decrease (Lemma 1).

- Suppose that the player becomes a pool member of other pools. By Lemma 2, their utility can be $P(s_{k+1}, c_{k+1})s_j/\beta$ at most, which is lower that their current utility by $P(s_j, c_j) - P(s_{k+1}, c_{k+1})$ (by Equation 5).

We now consider a player $j$ with rank higher than $k$. Again we show that none of the possible responses improves their utility. Notice first that by changing their allocation of stake, it can only hurt their utility since some of their stake ends up in pools with stake different than $\beta$, which can only lower their utility by Lemma 1. The other alternative is that the player becomes a pool leader. Since their rank is higher than $k$, the (non-myopic) stake of the pool contains only their own stake, which by Lemma 1 is again no better than the current utility.

We now sketch the full argument that considers the more complex strategies of possibly simultaneously delegating and creating a pool for each player (we remark that this case is also subsumed in the two-stage game in Section 5). Note that the desirability and thus the rank of the pools does not depend on the size of the pools. So if we allow strategies where a player is pool leader and simultaneously delegates some stake to other pools, then the perfect strategies remain Nash equilibria. In addition, it is easily verified that Lemmas 1, 2 hold also in this case.

- If a player $\in \{1, ..., k\}$ with stake $s$ and cost $c$ increases their margin from $m^*$ to $m'$ and delegates stake $s - \lambda$ to other pools then their pool will have rank higher than $k$ and their utility will become $\frac{\lambda}{\lambda} \cdot (r(\lambda, \lambda) - c) + P(s_{k+1}, c_{k+1}) \cdot \frac{s-\lambda}{\beta}$ which is no higher than $\frac{\lambda}{\beta} \cdot P(\lambda, c) + P(s_{k+1}, c_{k+1}) \cdot \frac{s-\lambda}{\beta}$ because $\frac{r(\sigma, \lambda) - c}{\sigma}$ increasing for $\sigma \leq 1/k$ (Lemmas 1). This is at most $(m^* + (1 - m^*) \cdot \frac{\lambda}{\beta}) \cdot P(s, c) + P(s_{k+1}, c_{k+1}) \cdot \frac{s-\lambda}{\beta}$ that is equal to their current utility.

- If a player $\in \{1, ..., k\}$ with stake $s$ and cost $c$ decreases their margin from $m^*$ to $m$ and simultaneously transfers stake $s - \lambda$ to other pools, then the desirability of their pool remains the same, increases or decreases. We will prove that in all cases their utility will be at most their current utility $(m^* + (1 - m^*) \cdot \frac{s}{\beta}) \cdot P(s, c)$.

  1. If the desirability of their pool remains the same, then (i) the utility for the part of their stake that remains in their pool denoted by $\lambda$ will decrease because of the lower margin or will remain the same and (ii) the utility for the stake that has been transferred to other pools denoted by $s - \lambda$ will also decrease because these pools have the same desirability and their non-myopic stake will become higher than $1/k$.

  2. If the desirability of their pool decreases, then the rank of their pool will become higher than $k$ regardless the stake this player delegated to other pools. So again the utility for both parts of stake will decrease.

  3. If the desirability of their pool increases then their utility will become $(m + (1 - m) \cdot \frac{\lambda}{\beta}) \cdot P(\lambda, c) + \frac{s-\lambda}{\beta} \cdot P(s_{k+1}, c_{k+1}) \leq (m^* + (1 - m^*) \cdot \frac{\lambda}{\beta}) \cdot P(s, c) + \frac{s-\lambda}{\beta} \cdot P(s_{k+1}, c_{k+1}) = (m^* + (1 - m^*) \cdot \frac{s}{\beta}) \cdot P(s, c)$.

- If a player $\in \{1, ..., k\}$ with stake $s$ and cost $c$ does not change margin and transfers stake $s - \lambda$ to other pools then again their utility will become $\frac{\lambda}{\lambda} \cdot (r(\lambda, \lambda) - c) + P(s_{k+1}, c_{k+1}) \cdot \frac{s-\lambda}{\beta}$ because their pool will have rank higher than $k$.

- If a player $\in \{k+1, ..., n\}$ with stake $s$ and cost $c$ creates a pool with stake $\lambda$ and delegates the remaining stake to other pools then their pool will have rank lower than $k$ so their utility will be $(r(\lambda, \lambda) - c) + \frac{s-\lambda}{\beta} \cdot P(s_{k+1}, c_{k+1}) \leq P(\lambda, c) \cdot \frac{\lambda}{\beta} + \frac{s-\lambda}{\beta} \cdot P(s_{k+1}, c_{k+1})$ which is not higher than their current utility $\frac{s}{\beta} \cdot P(s_{k+1}, c_{k+1})$.

□

It is interesting to note that in the first case of the proof of Theorem 4, the pool leader of a pool with stake $\beta$ decreases their margin. This does not affect our equilibrium argument since by the definition of non-myopic stake, the stake of their pool remains the same and hence the non-myopic utility is unaffected. But this pool will score a higher desirability and in the real world far-sighted pool members may prefer it and, in such case, its size may increase beyond $\beta$. This raises the question whether perfect strategies are stable when the players play non-myopically beyond the strict definition that is captured by the way we have considered so far in the analysis. To answer this question and understand the implications of these far-sighted strategies, we can conduct a "two-stage" game analysis which we present in Section 5.

### 4.3 Sybil resilience and whale stakeholder analysis

We now turn to the analysis of Sybil attacks as well as of the effect that large ("whale") stakeholders have in the game. Recall that in the previous section we restricted players to having stake at most $\beta = 1/k$ and each one creating at most one pool, hence explicitly excluding Sybil attacks and whale stakeholders. To remove these constraints, we consider an extended setting that involves a set of $\tilde{n} \leq n$ agents, each one with (private) stake $\tilde{s}_1, \ldots, \tilde{s}_{\tilde{n}}$ and associated (private) cost $\tilde{c}_1, \ldots, \tilde{c}_{\tilde{n}}$. Each agent $i$ can declare themselves as a single player in the stake-pool game as long as $\tilde{s}_i \leq \beta$, or alternatively declare more than one players (called *Sybils*) splitting their stake in some way between the declared players. This "pre-game" stage defines a specific instance of the stake-pool game. The utility of each agent is the sum of the utility of all the players that the agent controls.

We analyze two scenarios in this setting. In the first one, there is a utility non-maximizer agent with total stake less than $1/2$, who creates $k/2$ players, potentially lying about their costs, with the objective of dominating the system by creating $k/2$ saturated pools at the Nash equilibrium. In the second scenario, a utility maximizer agent creates $t > 1$ players that share their costs by using the same server. In both cases, to simplify the analysis, we will assume that the stake-pool game proceeds with players acting rationally and independently.

For a given agent, denote by $A \subseteq \{1, \ldots, n\}$ the set of players the agent introduces in the stakepool game. For each $A$, we denote by $(s_i^A, c_i^A)$ the stake and cost of the $i$-th player in the game, ordering them in decreasing order of potential profit, excluding $A$. Moreover, the maximum cost and the minimum stake, excluding players in $A$, will be denoted $c_{\max}^A$ and $s_{\min}^A$ respectively. We prove the following.

**Theorem 5.** *Consider an agent controlling a set of players $A$. First, if the agent has stake less than $\frac{k}{2} \cdot \left(s_{k/2+1}^A - \frac{c_{\max}^A}{R} \cdot (1 + \frac{1}{\alpha})\right)$ then it will control fewer than $k/2$ saturated pools at the Nash equilibrium, even if the agent is a utility non-maximizer. Second, if the agent is a utility maximizer with cost $\tilde{c}$ and stake less than $t \cdot \left(s_{k-t+1}^A - \frac{(c_{\max}^A - \tilde{c}/t)}{R} \cdot (1 + \frac{1}{\alpha})\right)$, it will control fewer than $t$ saturated pools at the Nash equilibrium for any $k \geq t > 1$.*

The proof is described in Appendix A.3. We observe that in both cases, the minimum stake needed by the Sybil attacker agent is asymptotically linear in the number of stake pools ($k/2$ in the first case and $t$ in the second). Moreover, the coefficient, in both cases, can be adjusted by varying the Sybil resilience parameter $\alpha$.

Specifically, when $\frac{c_{\max}^A}{R} < s_{\min}^A$, these bounds are positive for suitable value of $\alpha$; in particular, the higher $\alpha$ is, the higher these bounds become. Note that $s_{k/2+1}^A$ and $s_{k-t+1}^A$ are nondecreasing in $\alpha$, because the ordering of the remaining agents depends on $P(s_i, c_i)$ and thus also in $\alpha$ (the higher $\alpha$ is the higher impact agents' stake has on the ordering). For example, in the first case when $R = 1$ and $k = 10$, and the stake and cost are sampled from a Pareto distribution with parameter $\alpha = 2$ and the uniform distribution from $[0.0005, 0.0010]$ respectively, if we choose $\alpha = 0.5$ then $c_{k/2+1}^A = 0.00076024, s_{k/2+1}^A = 0.02002176$. Then if a non-utility attacker declares cost $c = 0.9 \cdot c_{k/2+1}^A$, the stake required for the attack is at least $0.0989$. This is not far from optimal, since the largest possible lower bound is $5 \cdot 0.02002176 = 0.1001088$, which would apply to the setting of negligible costs and a choice of $\alpha$ that goes to $+\infty$.

Next we examine the probability under reasonable probability distributions that there exists an agent who has stake more than $\frac{k}{2} \cdot s_{k/2+1}^A$, which allows them to engage in Sybil behavior in the above settings (i.e., with negligible costs and a choice of $\alpha$ that goes to $+\infty$).

Let $S_i$ and $s_i = \frac{S_i}{\sum_{i=1}^{\tilde{n}} S_i}$ be the absolute and the relative stake respectively of agent $i$. Let $S_1, ..., S_{\tilde{n}}$ be independent samples from random variable $X$ that follows the upper truncated Pareto distribution [10] with parameter $a \neq 0$. Let $\theta$ and $T$ be the minimum and maximum value of the distribution, respectively. Then the cumulative function of $X$ is $F_X(x) = \frac{1-(\frac{\theta}{x})^a}{1-(\frac{\theta}{T})^a}$ when $\theta \leq x \leq T$. Also if $X_r$ is the stake of the agent with the $r$-th smallest stake, then the cumulative function of $X_r$ is $F_{X_r}(x) = \sum_{j=r}^{\tilde{n}} [\binom{\tilde{n}}{j} \cdot F_X^j(x) \cdot (1-F_X(x))^{\tilde{n}-j}]$, see [8]. We also denote by $S_i = X_{\tilde{n}-i+1}$ the stake of the agent with the $i$-th highest stake. Let $f_{S_{\frac{k}{2}+1}}(t)$ be the density function of $S_{\frac{k}{2}+1}$ and $F_B(k; \tilde{n}, p) = \sum_{i=0}^{k} \binom{\tilde{n}}{i} \cdot p^i \cdot (1-p)^{\tilde{n}-i}$ the cumulative function of Binomial distribution. The following theorem quantifies the probability that a Sybil attack is possible.

**Theorem 6.** *Assume that $S_1, \ldots, S_{\tilde{n}}$, where $S_i$ is the absolute stake of agent $i$, are drawn from an upper truncated Pareto distribution with parameters $a, \theta, T$. Then when $\delta = \left( \frac{1-(\frac{\theta}{T})^a}{1-(\frac{\theta \cdot k}{2 \cdot T})^a} \right) \cdot (1 - \frac{k}{2\tilde{n}}) - 1 > 0$:*

$$Pr(s_1 > \frac{k}{2} \cdot s_{\frac{k}{2}+1}) \leq e^{-\delta^2 \mu / 3},$$

*where $\mu = \tilde{n} \cdot F_X(\frac{2T}{k})$.*

For the proof see Appendix A.4.

Note that if we take $a = 1$, $\frac{\theta}{T} = 1/100,000$ and $k = 100$, then in order for $\delta$ to be positive, it suffices $\tilde{n} > 150,000$, a reasonable number of users of a general cryptocurrency. Also if we choose higher $\tilde{n}$ or $\theta$ and lower $T$, then $\delta$ will increase. It holds that $\delta$ is

- increasing as a function of $\tilde{n}$ and decreasing as a function of $T$ and $a$

- increasing as a function of $k$ if and only if $\frac{\theta^a \cdot k^{a-1}}{2^a \cdot T^a} \cdot (k + 2 \cdot a \cdot \tilde{n} - a \cdot k) > 1$. In particular when $a = 1$, $\delta$ is increasing as a function of $k$ if and only if $T < \theta \cdot \tilde{n}$.

## 5 A Two-Stage Game Analysis

We will next prove that our reward sharing scheme effectively retains the same perfect equilibria outcome of Theorem 4 also in a more realistic two-stage or "inner-outer game." The advantages of this approach are as follows: (i) it allows us to analyze non-myopic moves in response to pool leaders changing margin or allocation, (ii) it allows us to remove the assumption that a player can be either a pool leader or a pool member, (iii) in this setting when a pool has not been activated, we define its desirability to be zero, something that gives us a more realistic result, because in practice only pools that have already been created will be ranked; (iv) in this game we break ties in ranking in arbitrary ways, not only according to potential profit. We note that similar non-myopic type of play has already been considered in other settings, notably in *Cournot Equilibria*, as is discussed in the introduction and related work.

Our "inner-outer game" consists of two games. In the *outer* game, player $i$ decides on the margin $m_i$ and on the stake $\lambda_i$ to be allocated to its own pool, in case the player will decide to activate it in the inner game. So a strategy of a player $i$ in the outer game is a tuple $(m_i, \lambda_i)$ of margin and allocated stake, and let $(\vec{m}, \vec{\lambda})$ be the joint strategy of the outer game. Each joint strategy of the outer game determines one inner game.

In the *inner* game, the margins $\vec{m}$ and the stakes $\vec{\lambda}$, which potential pool leaders would allocate to their pools, are given, and the strategies of the players are their allocations. So in the inner game determined by $(\vec{m}, \vec{\lambda})$, a strategy of player $i$ is $S_i^{(\vec{m},\vec{\lambda})} = \vec{a}_i$, and a joint strategy is $\vec{S}^{(\vec{m},\vec{\lambda})}$. Note that if a player $i$ decides to activate its own pool, which means $a_{i,i} > 0$, then the player is committed to allocate stake $\lambda_i$ to its own pool, where $\lambda_i$ is part of the strategy of the outer game. So $a_{i,i} \in \{0, \lambda_i\}$. We assume, that in the inner game the players decide their allocation with the goal of maximizing their non-myopic utility, as it is defined in 8. (Recall that we have assumed that each player can create at most one pool and that the utility an inactive pool gives to its members is zero.)

For a joint strategy $(\vec{m}, \vec{\lambda})$ of the outer game, we define the utility of a player $j$ to be equal to the non-myopic utility of this player in the equilibrium of the associated inner game. Formally $u_j^{\text{outer}}(\vec{m}, \vec{\lambda}) = u_j(\vec{S}^{(\vec{m}, \vec{\lambda})})$, where $\vec{S}^{(\vec{m}, \vec{\lambda})}$ is the unique equilibrium of the inner game determined by $(\vec{m}, \vec{\lambda})$. (We study also the case when the inner game has more than one or no equilibrium, by defining proper utilities and proper notion of equilibrium in this case, see Appendix 5).

In this framework, we describe a set of joint strategies that (i) are approximate non-myopic Nash equilibria of the outer game and (ii) have the characteristic that in the inner games defined by these joint strategies, all the equilibria form $k$ saturated pools. Recall that a pool is *saturated* when its stake is at least $\beta$. The pool leaders of these pools in these equilibria of the inner games are again the players with the highest values $P(s_i, c_i)$.

The intuition for how the set of margins of these joint strategies is determined is the following: The $k$ players with the highest values $P(s_i, c_i)$ set the maximum possible margin, as long as their pools belong to the $k$ most desirable pools (the pools with the highest desirability), no matter which margins the other players have currently. Note that if all players activated a pool of size $1/k$ with the same margin and their whole stake, then the $k$ pools with the highest potential profit ($P(s_i, c_i)$) would give the highest utility to their members. The formal analysis, the theorems and the proofs appear in Appendix 5.

**Definition of the game.** In order to also capture non-myopic moves in response to pool leaders changing margin or allocation, we define a two-stage game, the "inner-outer game". Similar non-myopic play has already been considered in other games, most notably in *Cournot Equilibria*, as is discussed in the introduction and related work. In this section we reuse *non-myopic utility* and *desirability* as defined in previous sections, but when a pool has not been activated in the inner game, we define its desirability to be zero. This gives us a more realistic result, because in practice only pools that have already been created will be ranked. In addition we remove the assumption that a player can be either a pool leader or a pool member. We order players by $P(s_i, c_i)$, and $i$ will denote the player with the $i_{th}$ highest value according to this ordering. We break ties in ranking in arbitrary ways, our analysis will hold for all of them. In fact, we define two games here, the *inner* game, which focuses on the allocation of stake, and the *outer* game, which focuses on the margins and on the stake that potential pool leaders commit to their pools. In the *outer* game, player $i$ decides on their margin $m_i$ and on how much stake $\lambda_i$ to allocate to their pool, should they decide to activate it in the inner game. So a strategy of a player $i$ in the outer game is a tuple $(m_i, \lambda_i)$ of margin and allocated stake. $(\vec{m}, \vec{\lambda})$ is a joint strategy of the outer game.

In the *inner* game, the margins $\vec{m}$ and the stakes $\vec{\lambda}$, that potential pool leaders would allocate to their pools, are given, and the strategies of the players are their allocations. So in the inner game determined by $(\vec{m}, \vec{\lambda})$, a strategy of player $i$ is $S_i^{(\vec{m}, \vec{\lambda})} = \vec{a}_i$, and a joint strategy is $\vec{S}^{(\vec{m}, \vec{\lambda})}$. Note that if a player $i$ decides to activate their own pool, which means $a_{i,i} > 0$, then they are committed to allocate stake $\lambda_i$ to their pool, where $\lambda_i$ is part of their strategy of the outer game. So $a_{i,i} \in \{0, \lambda_i\}$. We assume that players decide their allocation trying to maximize their non-myopic utility. Recall that we have assumed that each player can create at most one pool and that the utility that an inactive pool gives to its members is zero. Note that each joint strategy of the outer game determines one inner game.

## 5.1 Definition of Equilibria for Inner and Outer Game

**Definition 11.** A joint strategy $\vec{S}^{(\vec{m}, \vec{\lambda})}$ is a Nash equilibrium of the inner game defined by $(\vec{m}, \vec{\lambda})$ when for every player $j$

$$u_j(S_j'^{(\vec{m}, \vec{\lambda})}, \vec{S}_{-j}^{(\vec{m}, \vec{\lambda})}) \leq u_j(\vec{S}^{(\vec{m}, \vec{\lambda})}) \tag{6}$$

for every $S_j'^{(\vec{m}, \vec{\lambda})} \neq S_j^{(\vec{m}, \vec{\lambda})}$. This is the standard Nash equilibrium notion when the players try to maximize their non-myopic utility.

To define the non-myopic equilibrium of the outer game, let us temporarily assume that there is a *unique Nash equilibrium in every inner game*. Then we define the utility of player $j$ in the outer game, where players have selected joint strategy $(\vec{m}, \vec{\lambda})$, as: $u_j^{\text{outer}}(\vec{m}, \vec{\lambda}) = u_j(\vec{S}^{(\vec{m}, \vec{\lambda})})$, where $\vec{S}^{(\vec{m}, \vec{\lambda})}$ is the unique equilibrium of the inner game determined by $(\vec{m}, \vec{\lambda})$. So a joint strategy $(\vec{m}, \vec{\lambda})$ is an approximate $\epsilon$-non-myopic Nash equilibrium of the outer game when for every player $j$

$$u_j^{\text{outer}}(m_j', \vec{m}_{-j}, \lambda_j', \vec{\lambda}_{-j}) \leq u_j^{\text{outer}}(\vec{m}, \vec{\lambda}) + \epsilon \tag{7}$$

for every $(m_j', \lambda_j') \neq (m_j, \lambda_j)$.

When there are multiple equilibria in the inner game, we define $u_j^{\text{outer}}(\vec{m}, \vec{\lambda})$ as *the set* of values $u_j(\vec{S}^{(\vec{m}, \vec{\lambda})})$, where $\vec{S}^{(\vec{m}, \vec{\lambda})}$ is a Nash equilibrium of the inner game determined by $(\vec{m}, \vec{\lambda})$.

Let

$$u_j^{\text{outer,up}}(\vec{m}, \vec{\lambda}) = \begin{cases} \sup u_j^{\text{outer}}(\vec{m}, \vec{\lambda}) & \text{if } u_j^{\text{outer}}(\vec{m}, \vec{\lambda}) \neq \emptyset, \\ -\infty & \text{elsewhere.} \end{cases} \tag{8}$$

In the same way we define:

$$u_j^{\text{outer,low}}(\vec{m}, \vec{\lambda}) = \begin{cases} \inf u_j^{\text{outer}}(\vec{m}, \vec{\lambda}) & \text{if } u_j^{\text{outer}}(\vec{m}, \vec{\lambda}) \neq \emptyset, \\ -\infty & \text{elsewhere.} \end{cases} \tag{9}$$

Note that when $u_j^{\text{outer}}(\vec{m}, \vec{\lambda})$ is not empty, it is a non-empty bounded subset of the reals and therefore always has both supremum and infimum: Upper- and lower bounds are given by $R$ and $(-\max\{c_1, \ldots, c_n\})$ respectively.

**Definition 12.** A joint strategy $(\vec{m}, \vec{\lambda})$ is an $\epsilon$-non-myopic Nash equilibrium when for every player $j$

$$u_j^{\text{outer,up}}(m_j', \vec{m}_{-j}, \lambda_j', \vec{\lambda}_{-j}) \leq u_j^{\text{outer,low}}(\vec{m}, \vec{\lambda}) + \epsilon \tag{10}$$

for every $(m_j', \lambda_j') \neq (m_j, \lambda_j)$.

For the formal theorems and proofs referring to the two-stage game see in Appendix A.5.

## 6 Deployment Considerations

In this section we overview various deployment considerations of our RSS solution as well as we address specific attacks and deviations against our reward sharing scheme, specifically, (i) pools that underperform in general, (ii) participants who play myopically, (iii) pools that censor undesirable delegation transactions, (iv) pool leaders not truthfully declaring their costs, and (v) parties who try to gain advantage by exploiting how wealth may compound over time ("the rich get richer" problem) in a series of iterations of the game.

Regarding deployment, in order to facilitate the use of an RSS within a PoS cryptocurrency, e.g., [30, 25, 13, 6], the ledger should be enhanced to enable special transactions which allow players to delegate their stake to a pool and reassign it at will during the course of the execution. Describing in more detail the exact cryptographic mechanism for performing this operation is outside the scope of the present paper. It is sufficient to note that the mechanism is simple and very similar to issuing public-key certificates; see e.g., [25] for a description of such a delegation mechanism. Recall that in a PoS cryptocurrency, the protocol is executed by electing participants in some way based on the stake they possessed in the ledger; informally every protocol message is signed on behalf of particular coin that is verifiably elected for that particular point of the protocol's execution. In the stake pool setting, the PoS protocol will be executed with the pool leaders representing the pool members whenever the coin of a member is elected for protocol participation.

**Ill-performing stake pools.** In our system, rewards for a pool are calculated based on the declared stake of the pool leader as well as the stake delegated to that pool. This provides an opportunity for a pool leader to declare a competitive pool and subsequently do not provide the service that it promised (presumably gaining in terms of the actual cost that system maintenance incurs). This can be addressed by calibrating the total rewards $R$ to depend also on the total performance of the system as evidenced in the distributed ledger. For instance, in a PoS blockchain, it is possible to count the number of blocks that were produced in a period of time and compare that value to its expectation. In case the actual number of blocks is below expectation we may reduce $R$ accordingly (effectively punishing all pools) and in this way generating a counter-incentive to deviate from system maintenance according to the protocol. Note that punishing all pools in case of underperformance makes sense due to the possibility of mining games [16, 24] which may be used by pools to attack each other in case we use performance as indicator for punishment. However, punishing everyone may be hard to parameterise as a large reduction in $R$ will be unfair to genuinely performing participants (who will be losing rewards due to the ill performance of others) while a small reduction may be insufficient as a counterincentive. If the underlying blockchain is also "fair" (in the sense of [35]) then it might be also possible to penalise only specific pools that underperform and hence be able to better fine tune performance sensitivity. It is an interesting question to design such robust performance metrics that can be used in the context of a reward sharing scheme.

**Players who play myopically and Rational Ignorance.** Myopic play is not in line with the way we model rational behavior in our analysis. We explain here how it is possible to force rational parties to play non-myopically. With respect to pool leaders we already mentioned in Section 2.3 that rational play cannot be myopic since the latter leads to unstable configurations with unrealistically high margins that are not competitive. Next we argue that it is also possible to force pool members to play non-myopically. The key idea is that the effect of delegation transactions should be considered only in regular intervals (as opposed to be effective immediately) and in a certain restricted fashion. This can be achieved by e.g., restricting delegation instructions to a specific subset of stakeholders at any given time in the ledger operation and making them effective at some designated future time of the ledger's operation. Due to these restrictions, players will be forced to think ahead about the play of the other players, i.e., stakeholders will have to play based on their understanding of how other stakeholders will as well as the eventual size of the pools that are declared. A related problem is that of rational ignorance, where there is some significant inertia in terms of stakeholders engaging with the system resulting to a large amount of stake remaining undelegated. This can be handled by calibrating the total rewards $R$ to lessen according to the total active stake delegated, in this way incentivising parties to engage with the system.

**Censorship of delegation transactions.** In this attack, a pool (or a group of pools) censors delegation transactions that attempt to re-delegate stake or create a new pool that is competitive to the existing ones. In the extreme version of this attack a "cartel" of pool leaders control the whole PoS ecosystem and prevent new (potentially more competitive) pools from entering or existing members from delegating their stake. Actually, this is a typical threat to all "political" systems in which power is delegated to representatives. However, in PoS systems even a single pool that does not censor attacks is sufficient to prevent this attack assuming there is sufficient bandwidth to record the delegation transactions in the blocks that are contributed by that pool. It is an interesting question to address the case where all stake pools form a coalition that decides to prevent any more pools from being created. A potential way forward to preventing such abuse of power by pool leaders, is by either creating the right system safeguards and incentives for the coalition to break or rely on direct member participation that will override the pool leader cartel. In this latter case, pool members acting as system "watchdogs", without getting any reward, could still create alternative blocks, that take precedence over the blocks issued by the block leader in this way creating a ledger fork along which censorship is stopped.

**Costs and incentive compatibility.** In our analysis, we assumed for simplicity that the costs are publicly known; in reality the actual costs for participating in the collaborative project are known only by the player, who may lie about it in the cost declaration. This will happen when the players may see it as an advantage to lie about their cost. This problem is one of mechanism design which has objective to design an incentive compatible mechanism, i.e., a mechanism that gives incentives to players to declare their costs truthfully. We next argue that, in fact, our RSS is incentive-compatible as presented. Let us consider the perfect Nash equilibrium from Definition 9 in which the utilities are given by Equation 5. Suppose that a pool leader $j$ declared a different cost $\hat{c}_j$, but remained pool leader. Since $P(s_j, \hat{c}_j) - P(s_j, c_j) = c_j - \hat{c}_j$, the player will not get any benefit from lying. To see this, let $u_j(\hat{c}_j | c_j)$ denote the utility when the player declares cost $\hat{c}_j$ instead of the true cost $c_j$. Then by taking into account the cost, we have $u_j(\hat{c}_j | \hat{c}_j) = u_j(\hat{c}_j | c_j) - c_j + \hat{c}_j$. Also from Equation 5, we see that $u_j(c_j | c_j) - u_j(\hat{c}_j | \hat{c}_j) = P(s_j, c_j) - P(s_j, \hat{c}_j)$. Putting them together we see that $u_j(\hat{c}_j | c_j) = u_j(c_j | c_j)$, thus the player has no reason to lie. With similar reasoning, a pool leader has no reason to lie by raising his cost so much that the rank of his pool increases above $k$. Similar considerations, show that no pool member (i.e., a player whose pool, if created, would have had a rank at least $k+1$) has an incentive to lie. This includes the special case of the player with rank $k+1$. As a conclusion, we see that under the assumption that the players end up at a perfect equilibrium, it is a dominant strategy to declare the true cost. As a side note, we could also adapt any similar reward scheme to implement the Vickrey-Clarke-Groves (VCG) mechanism, cf. [31], which applies to all mechanism design problems. In this particular case, the VCG mechanism, would ask the players to declare their costs $\vec{c}$, but the reward scheme would *use a different vector of costs $\vec{c}'$ for the game*. The new costs $\vec{c}'$ will be such that the desirability of the player with rank $j \leq k$ would be slightly superior to the desirability of the player with rank $k+1$.

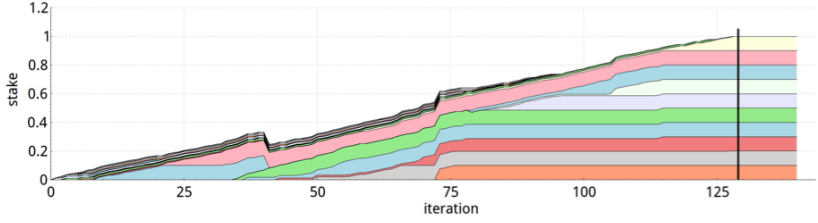

Figure 6: Example dynamics of our reward sharing scheme ($c \in [0.001, 0.002]$, $\alpha = 0.02$) showing convergence to decentralization.

**"Rich getting richer" considerations.** In a PoS deployment, our game will be played in epochs with each iteration succeeding the previous one. Using the mechanisms we described above regarding censorship and Sybil resilience, it is easy to see that players are not bound by their past decisions and thus they will treat each epoch as a new independent game. A special consideration here is what frequently is referred to as the "rich get richer" problem, i.e., the setting where the richest stakeholder(s) amass over time even more wealth due to receiving rewards leading to an inherently centralised system (it is sometimes believed that this issue is intrinsic to only PoS systems but in fact it equally applies to PoW systems, cf. [23]). In order to address this issue we observe that the maximum rewards obtained by each pool at each epoch are in the range $[R/(1+\alpha)k, R/k]$ with $\alpha \in [0, +\infty)$ determining the size of the range which controls how much more rich pools (i.e., pools with rich pool leaders who can pledge more stake to their pool) benefit. It follows that using $\alpha$ we can control the disparity created by the reward mechanism by choosing $\alpha$ closer to 0, with the choice $\alpha \to 0$ achieving a perfectly "egalitarian" effect where rich pools and poor pools of the same size are receiving exactly the same rewards, something that does not affect the relative stake from epoch to epoch if we do not take into account margins. Note that while this completely equalises the power of a "rich dollar" versus a "poor dollar" (cf. [23]) it comes with the downside of a reduction of the system's resilience to Sybil attacks. Given we have no way of guaranteeing the independence of the players as declared in the stake pool game, we offer a tradeoff between egalitarianism and Sybil resilience.

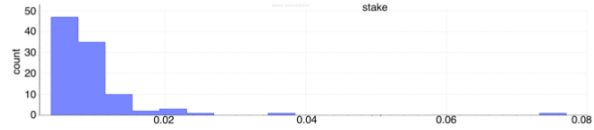

Figure 7: The stake-distribution used for all experiments (but see the paragraph on other choices of the parameter at the end of this appendix 7).

## 7 Experimental Results

In this section we describe how the experiments were executed.

**Initialization**

We simulate 100 players, and we use $k = 10$ for the desired number of pools. We assign stake to each player by sampling from a *Pareto distribution*[7] with parameter 2, truncated to ensure that no player has higher stake than $\frac{1}{k}$. The distribution is shown in Figure 7.

Furthermore, we assign a *cost* to each player, uniformly sampled from $[c_{\min}, c_{\max}]$, where both $c_{\min}$ and $c_{\max}$ are configurable.

**Player strategies**

The following *strategies* are available for players:

- A player can *lead a pool* with *margin* $m \in [0, 1]$.

- Alternatively, a player can *delegate* their stake to existing pools. They can freely choose how much stake to delegate to each pool, and they do not have to delegate all of their stake.

Initially, there are no pools and all players play the second strategy and do not delegate any stake. When it is a player's turn to *move*, they can freely switch to another strategy:

- A pool leader can keep their pool, but change their margin, or close their pool and delegate to other pools.

- A player without pool can change its delegation or start a pool.

If a pool leader decides to close their pool, all stake delegated to that pool by other players automatically becomes un-delegated.

**Simulation step**

In each step, we look for a player with a move that increases the player's utility by a minimal amount[8]. If a player with such a move is found, we apply that move and repeat. If not, we have reached an equilibrium. We have to deal with the technical problem that for each player, there is an infinity of potential moves to consider. We solve this problem in an approximate manner as follows:

- For pool moves, instead of considering all margins in $[0, 1]$, we restrict ourselves to one or two margins, namely 1 (to consider the case where the player plans running a one-man pool) and the highest margin $m < 1$ that has a chance (we make this precise below) to attract members (calculated to a precision of $10^{-12}$ if such a margin exists).

---

[7]Distributions of wealth tend to follow such a distribution, see [10], [34] and https://en.wikipedia.org/wiki/Pareto_distribution.

[8]Specifically, we consider utility $u_{\text{new}}$ non-trivially better than utility $u_{\text{old}}$ iff $u_{\text{new}} > u_{\text{old}} + 10^{-8}$.

- For delegation moves, we approximate the optimal delegation strategy using a local search heuristic ("beam search"[9]). Furthermore, we restrict ourselves to a resolution of multiples of $10^{-8}$ of player stake.

**How players choose their strategy in a non-myopic way**

We have the problem of how to avoid "myopic" margin increases: It is always tempting for a pool leader to increase their margin (or for a delegating player to start a pool with a high margin), but such a move only makes sense if sufficiently few other players have incentive to create more desirable pools during the next steps. To be more precise: If a player $A$ contemplates running a pool with margin $m < 1$, $A$ wants players to delegate to their pool and saturate it (if they wanted to run a one-man pool instead, the margin would be irrelevant and could be set to 1). Only pools with rank $\leq k$ attract delegations and have a chance of becoming saturated, so running a pool with margin $m$ only makes sense if the pool can reasonably be expected to end up with rank $\leq k$. This means that when running such a pool, at most $k-1$ other players should have incentive to run more desirable pools.

In order to determine whether $m$ satisfies this condition, we look at all other players. For players who already run pools, we assume that they will continue running their pools and keep their margins. For each other player $B$, we check whether there exists a margin $m'$ such that by creating a pool with margin $m'$ and by assuming that that pool would have rank $m' \leq k$, $B$ would increase its utility. Let $B$ have stake $s$, costs $c$ and utility $u$. If $B$ manages to create a pool with rank $m' \leq k$, then that pool's stake will be $\sigma := \max(s, \beta)$, and we can calculate its rewards $r$. Setting $q := \frac{s}{\sigma}$ and plugging in pool leader utility, we are looking for the minimal margin $m'$ satisfying

$$(r - c)\left[m' + (1 - m')q\right] > u.$$

We see that $r > c$ is a necessary condition. For $q = 1$ (i.e. $s \geq \beta$), $m' = 0$ is the obvious solution. For $q < 1$, we get

$$m' > \frac{u - (r - c)q}{(r - c)(1 - q)},$$

and we pick $\frac{u-(r-c)q}{(r-c)(1-q)}$ as margin for player $B$. We end up with a list of pools, one for each player, and we only allow $A$ to consider their pool move with margin $m$ if $A$'s pool would be amongst the $k$ most desirable pools in this list.

Note that this procedure of choosing the strategy represents the fact that players in our theoretical analysis try to maximize their non-myopic utility.

**Explaining the results**

The outcome of each simulation is a diagram with various plots, visualizing the dynamics, and a table with data describing the reached equilibrium. For the simulations reported here, we have always used the same stake distribution (sampled randomly from a Pareto distribution, as explained above) to make results more comparable (see Figure 7).

**dynamics** displays the dynamic assignment of stake to pools. At the end of each simulation, once an equilibrium has been reached, we expect all stake to be assigned to ten pools of equal size.

**pools** shows the number of pools over time — this should end up at ten pools.

In the tables describing the equilibrium (all found in Appendix 7), the meaning of the columns is as follows:

**player** Number of the player who leads the pool. Players are ordered by their *potential* $P(s, c)$ (cost $c$, stake $s$). Our expectation is to end up with ten pools, led by players 1–10.

---

[9] See https://en.wikipedia.org/wiki/Beam_search.

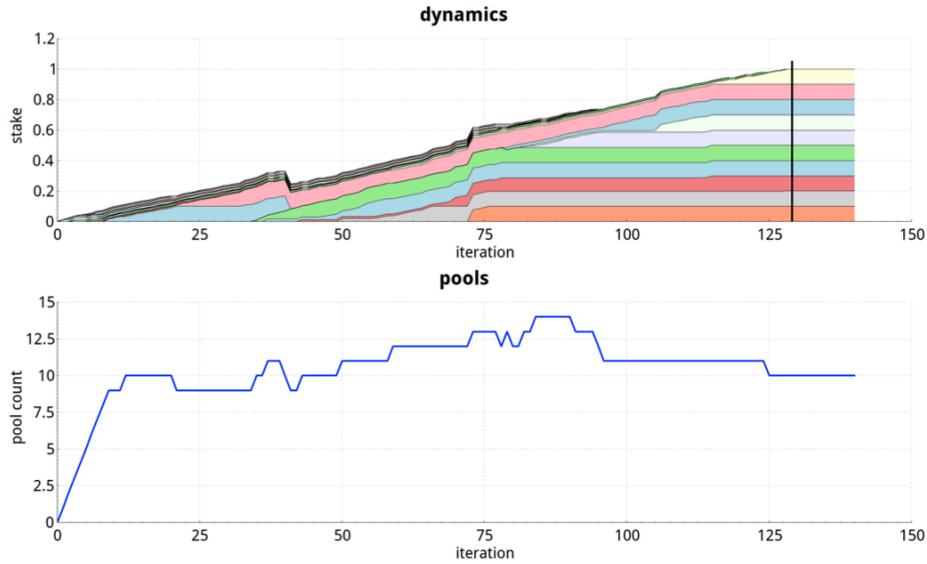

Figure 8: Low costs, low stake influence ($c \in [0.001, 0.002]$, $\alpha = 0.02$).

**rk** Pool *rank*. We expect our final pools to have ranks 1–10.

**crk** Pool leader's *cost-rank*: The player with the lowest costs has cost-rank 1, the player with the second lowest costs has cost-rank 2 and so on. For low values of $\alpha$, this should be close to the pool rank.

**srk** Pool leader's *stake-rank*: The player with the highest stake has stake-rank 1, the player with the second highest stake has stake-rank 2 and so on. For high values of $\alpha$, this should be close to the pool rank.

**cost** Pool costs.

**margin** Pool margin.

**player stake** Pool leader's stake.

**pool stake** Pool stake (including leader and members).

**reward** Pool rewards (before distributing them amongst leader and members).

**desirability** Pool desirability.

We show the results of six exemplary simulations with various costs and values for parameter $\alpha$ (which governs the influence of pool leader stake on pool desirability):

- low costs and low $\alpha$, see figure 8 and table 1.
- lows costs and medium $\alpha$, see figure 9 and table 2.
- lows costs and high $\alpha$, see figure 10 and table 3.
- high costs and low $\alpha$, see figure 11 and table 4.
- high costs and medium $\alpha$, see figure 12 and table 5.
- high costs and high $\alpha$, see figure 13 and table 6.

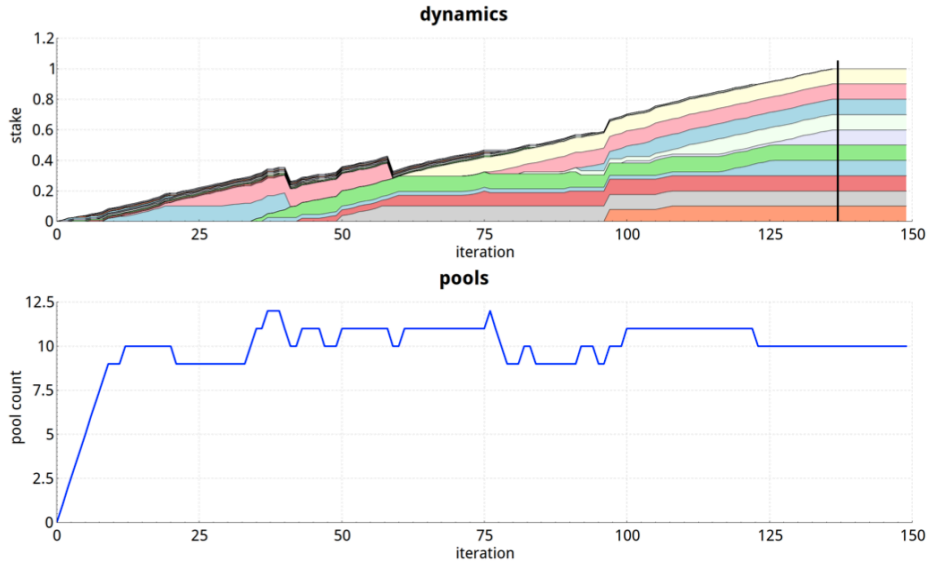

Figure 9: Low costs, medium stake influence ($c \in [0.001, 0.002]$, $\alpha = 0.05$).

**Additional experiments allowing simultaneous moves**

As explained above, in each simulation step we look for *one* player with an advantageous move and allow that player to make his move. In a real-world blockchain system however, players will probably be allowed to move concurrently, so we did some additional experiments allowing for this. Instead of picking just one player, we allowed several players with utility-increasing moves to make their move in one step. It is possible that such moves contradict each other (for example when one player closes a pool that a second player wants to delegate to). We handled this by applying the moves in order and dropping those that were invalid. Furthermore, in order to allow the system to stabilize, we blocked players from making "pool moves" (creating or closing a pool or changing the margin) too often by only allowing delegation moves for a number of steps after a player has made a pool move. Of course before we declare an equilibrium having been reached, we wait long enough to see whether any player wants to make a pool move after his waiting period is over. An example for five players being allowed to move simultaneously and a waiting period for the next pool move of 100 steps can be seen in figure 14.

**Other choices for the parameter of the Pareto distribution**

In all experiments discussed until now we used the same stake distribution of players drawn from a Pareto distribution with parameter 2 (shown in Figure 7). We picked this parameter for resulting in an apparently realistic distribution, but our results are not sensitive to this choice. To demonstrate this, we reran some of the experiments (for high costs and high $\alpha$) for different parameter values.

- Parameter 1.1, see figures 15, 16 and table 7.
- Parameter 1.3, see figures 17, 18 and table 8.
- Parameter 1.5, see figures 19, 20 and table 9.

# 8 Conclusions and Future Directions

We studied the design of reward sharing schemes (RSS) for collaborative projects, such as maintaining a PoS cryptocurrency, that promote decentralisation. Our main result is the design of an RSS that exhibits an equilibrium that provides a desired level of decentralisation in conjunction to an efficiency/Sybil-resilience tradeoff that can be calibrated via suitable parameter selection. We describe how our RSS can be deployed in the context of a PoS system such as Ouroboros [25] addressing a number of attacks and possible protocol deviations.

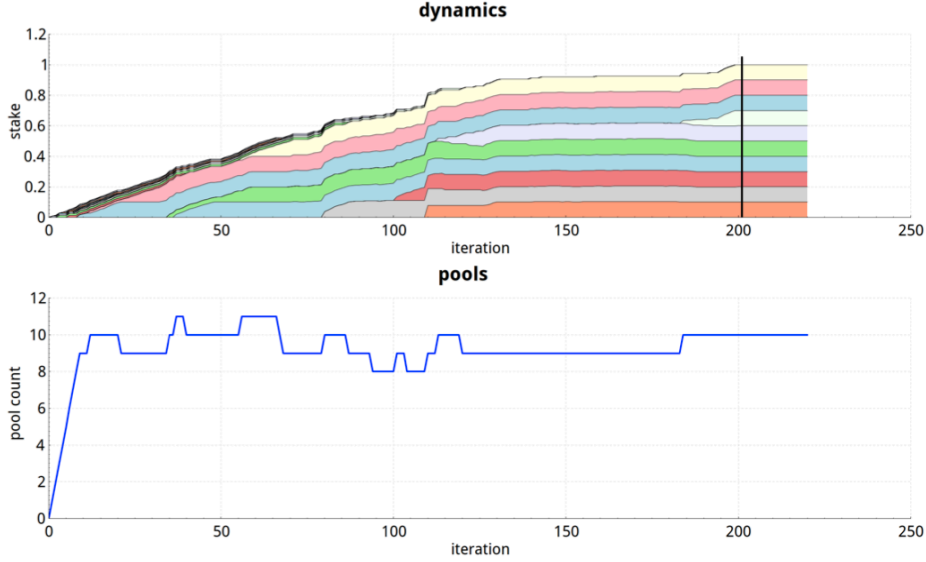

Figure 10: Low costs, high stake influence ($c \in [0.001, 0.002]$, $\alpha = 0.5$).

Table 1: Low costs, low stake influence ($c \in [0.001, 0.002]$, $\alpha = 0.02$).

| player | rk | crk | srk | cost | margin | player stake | pool stake | reward | desirability |
|---|---|---|---|---|---|---|---|---|---|
| 1 | 4 | 54 | 1 | 0.00156856 | 0.00898774 | 0.07704926 | 0.10000000 | 0.10154099 | 0.099073896587544 |
| 2 | 5 | 19 | 5 | 0.00121229 | 0.00125302 | 0.02052438 | 0.10000000 | 0.10041049 | 0.099073896587539 |
| 3 | 9 | 5 | 17 | 0.00108188 | 0.00088317 | 0.01216771 | 0.10000000 | 0.10024335 | 0.099073896587511 |
| 4 | 3 | 16 | 7 | 0.00120205 | 0.00063505 | 0.01694531 | 0.10000000 | 0.10033891 | 0.099073896587553 |
| 5 | 2 | 6 | 26 | 0.00108805 | 0.00053598 | 0.01075376 | 0.10000000 | 0.10021508 | 0.099073896587558 |
| 6 | 1 | 1 | 81 | 0.00100213 | 0.00047005 | 0.00613080 | 0.10000000 | 0.10012262 | 0.099073896587589 |
| 7 | 7 | 3 | 39 | 0.00105867 | 0.00047469 | 0.00898080 | 0.10000000 | 0.10017962 | 0.099073896587522 |
| 8 | 6 | 18 | 8 | 0.00121088 | 0.00042690 | 0.01635433 | 0.10000000 | 0.10032709 | 0.099073896587534 |
| 9 | 8 | 2 | 62 | 0.00103849 | 0.00026601 | 0.00693720 | 0.10000000 | 0.10013874 | 0.099073896587515 |
| 10 | 10 | 12 | 16 | 0.00115913 | 0.00011986 | 0.01224503 | 0.10000000 | 0.10024490 | 0.099073896587504 |

We view our work as a first step in the direction of understanding the design of mechanisms that promote decentralisation in a setting where multiple anonymous rational stakeholders operate and are interested in the maintenance of a collaborative project. There is a number of questions about reward schemes in this setting that are still open and worthy of further research which we hope our work will motivate. A first question is proving the uniqueness of equilibria with good properties and demonstrating rigorously, via a mathematical analysis, that such equilibria are efficiently reachable confirming the dynamics we have observed experimentally. Furthermore, it is interesting to investigate RSS's with a more general reward structure. In our work, we have considered reward schemes in which the reward of each pool is divided among its members proportionally to their stake and that pool operators perform all the work benefiting from their declared profit margin. An open research direction is to consider other organizational schemes in which the work and also the rewards of a pool are divided based on other factors, such as (weighted) Shapley value. With respect to Sibyl attacks and the "rich get richer" problem, in our scheme, increasing Sibyl-resilience affects inverse proportionally egalitarianism. An interesting research direction is to formally characterize the relationship between these two properties across the whole design space. Finally with respect to censorship, it is interesting to consider extended game theoretic models where censoring transactions is part of the strategic play

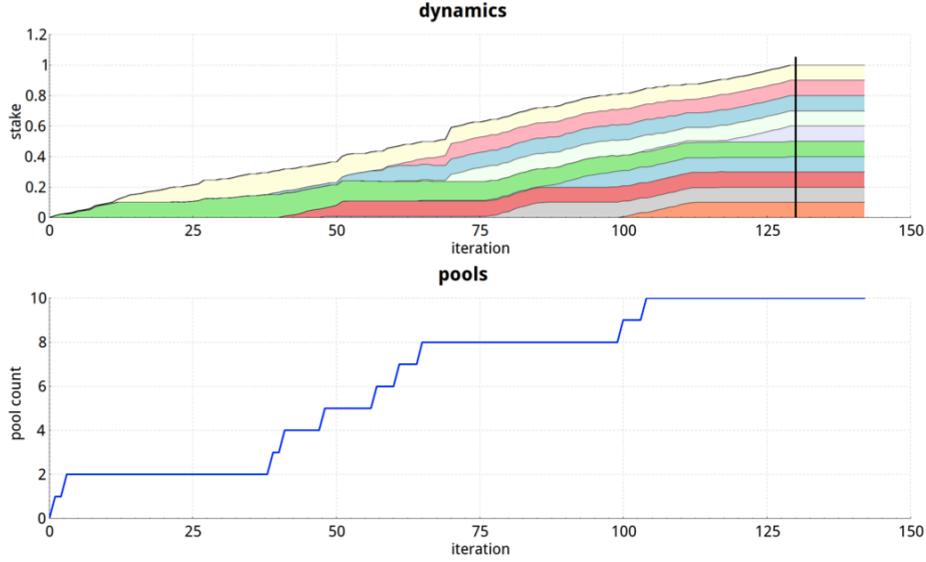

Figure 11: High costs, low stake influence ($c \in [0.05, 0.1]$, $\alpha = 0.02$).

Table 2: Low costs, medium stake influence ($c \in [0.001, 0.002]$, $\alpha = 0.05$).

| player | rk | crk | srk | cost | margin | player stake | pool stake | reward | desirability |
|---:|---:|---:|---:|---:|---:|---:|---:|---:|---:|
| 1 | 3 | 54 | 1 | 0.00156856 | 0.02828919 | 0.07704926 | 0.10000000 | 0.10385246 | 0.099390372362847 |
| 2 | 4 | 73 | 2 | 0.00173639 | 0.00742961 | 0.03741446 | 0.10000000 | 0.10187072 | 0.099390372362843 |
| 3 | 7 | 19 | 5 | 0.00121229 | 0.00424342 | 0.02052438 | 0.10000000 | 0.10102622 | 0.099390372362829 |
| 4 | 9 | 52 | 3 | 0.00156655 | 0.00297511 | 0.02507001 | 0.10000000 | 0.10125350 | 0.099390372362814 |
| 5 | 2 | 16 | 7 | 0.00120205 | 0.00255748 | 0.01694531 | 0.10000000 | 0.10084727 | 0.099390372362863 |
| 6 | 10 | 18 | 8 | 0.00121088 | 0.00217321 | 0.01635433 | 0.10000000 | 0.10081772 | 0.099390372362800 |
| 7 | 8 | 5 | 17 | 0.00108188 | 0.00136779 | 0.01216771 | 0.10000000 | 0.10060839 | 0.099390372362825 |
| 8 | 1 | 45 | 6 | 0.00152048 | 0.00090705 | 0.02002176 | 0.10000000 | 0.10100109 | 0.099390372362877 |
| 9 | 5 | 6 | 26 | 0.00108805 | 0.00059595 | 0.01075376 | 0.10000000 | 0.10053769 | 0.099390372362835 |
| 10 | 6 | 12 | 16 | 0.00115913 | 0.00063097 | 0.01224503 | 0.10000000 | 0.10061225 | 0.099390372362834 |

of pool leaders and pool members are on watch to protect the system (cf. Section 6).

# 9 Acknowledgements

The authors would like to thank Duncan Coutts for extensive discussions and many helpful suggestions. The second author was partially supported by H2020 project PRIVILEDGE # 780477. The third author was partially supported by the ERC Advanced Grant 321171 (ALGAME).

# References

[1] Nick Arnosti and S. Matthew Weinberg. Bitcoin: A natural oligopoly. *CoRR*, abs/1811.08572, 2018.

[2] K. Baclawski. *Introduction to Probability with R*. Chapman & Hall/CRC Texts in Statistical Science. CRC Press, 2008.

[3] Umang Bhaskar, Lisa Fleischer, Darrel Hoy, and Cien-Chung Huang. Equilibria of Atomic Flow Games are not Unique. *SODA*, pages 748–757, 2009.

[4] Umang Bhaskar and Phani Raj Lolakapuri. Equilibrium computation in atomic splittable routing games with convex cost functions. *arXiv preprint arXiv:1804.10044*, 2018.

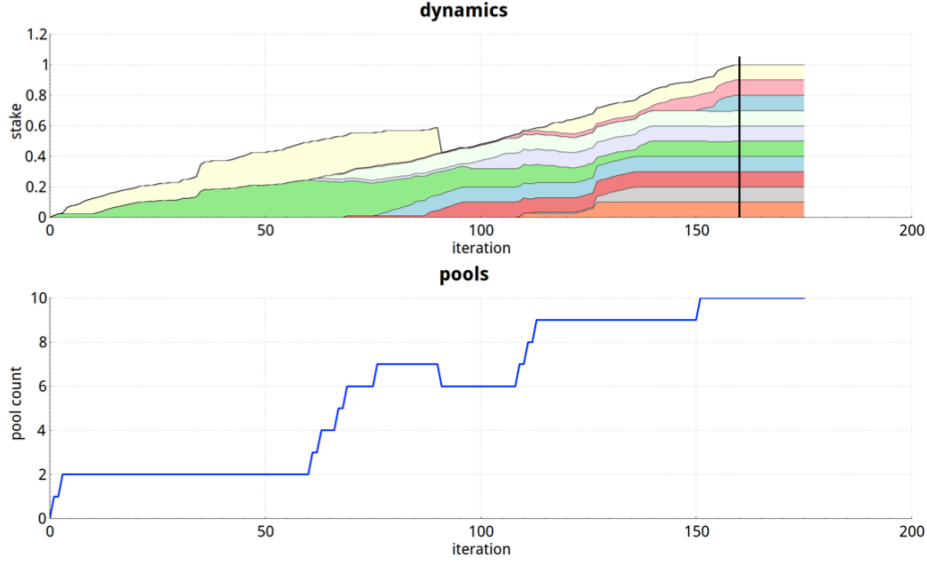

Figure 12: High costs, medium stake influence ($c \in [0.05, 0.1]$, $\alpha = 0.05$).

Table 3: Low costs, high stake influence ($c \in [0.001, 0.002]$, $\alpha = 0.5$).

| player | rk | crk | srk | cost | margin | player stake | pool stake | reward | desirability |
|---:|---:|---:|---:|---:|---:|---:|---:|---:|---:|
| 1 | 8 | 54 | 1 | 0.00156856 | 0.23109808 | 0.07704926 | 0.10000000 | 0.13852463 | 0.105305784121818 |
| 2 | 1 | 73 | 2 | 0.00173639 | 0.09972617 | 0.03741446 | 0.10000000 | 0.11870723 | 0.105305784121865 |
| 3 | 3 | 52 | 3 | 0.00156655 | 0.05102956 | 0.02507001 | 0.10000000 | 0.11253500 | 0.105305784121842 |
| 4 | 2 | 19 | 5 | 0.00121229 | 0.03433392 | 0.02052438 | 0.10000000 | 0.11026219 | 0.105305784121863 |
| 5 | 7 | 80 | 4 | 0.00181011 | 0.03273015 | 0.02135839 | 0.10000000 | 0.11067919 | 0.105305784121826 |
| 6 | 6 | 45 | 6 | 0.00152048 | 0.02935389 | 0.02002176 | 0.10000000 | 0.11001088 | 0.105305784121833 |
| 7 | 10 | 16 | 7 | 0.00120205 | 0.01831648 | 0.01694531 | 0.10000000 | 0.10847266 | 0.105305784121808 |
| 8 | 4 | 18 | 8 | 0.00121088 | 0.01552360 | 0.01635433 | 0.10000000 | 0.10817716 | 0.105305784121835 |
| 9 | 5 | 64 | 10 | 0.00163171 | 0.00394146 | 0.01470840 | 0.10000000 | 0.10735420 | 0.105305784121835 |
| 10 | 9 | 82 | 9 | 0.00181572 | 0.00298747 | 0.01487410 | 0.10000000 | 0.10743705 | 0.105305784121813 |

[5] Lars Brünjes, Aggelos Kiayias, Elias Koutsoupias, and Aikaterini-Panagiota Stouka. Reward sharing schemes for stake pools. *CoRR*, abs/1807.11218, 2018.

[6] Vitalik Buterin and Virgil Griffith. Casper the friendly finality gadget. *CoRR*, abs/1710.09437, 2017.

[7] Miles Carlsten, Harry A. Kalodner, S. Matthew Weinberg, and Arvind Narayanan. On the instability of bitcoin without the block reward. In Edgar R. Weippl, Stefan Katzenbeisser, Christopher Kruegel, Andrew C. Myers, and Shai Halevi, editors, *Proceedings of the 2016 ACM SIGSAC Conference on Computer and Communications Security, Vienna, Austria, October 24-28, 2016*, pages 154–167. ACM, 2016.

[8] G. Casella and R.L. Berger. *Statistical Inference*. Duxbury advanced series in statistics and decision sciences. Thomson Learning, 2002.

[9] Xi Chen, Christos H. Papadimitriou, and Tim Roughgarden. An axiomatic approach to block rewards. In *Proceedings of the 1st ACM Conference on Advances in Financial Technologies, AFT 2019, Zurich, Switzerland, October 21-23, 2019*, pages 124–131. ACM, 2019.

[10] David R. Clark. A note on the upper-truncated pareto distribution. *2013 Enterprise Risk Management Symposium*.

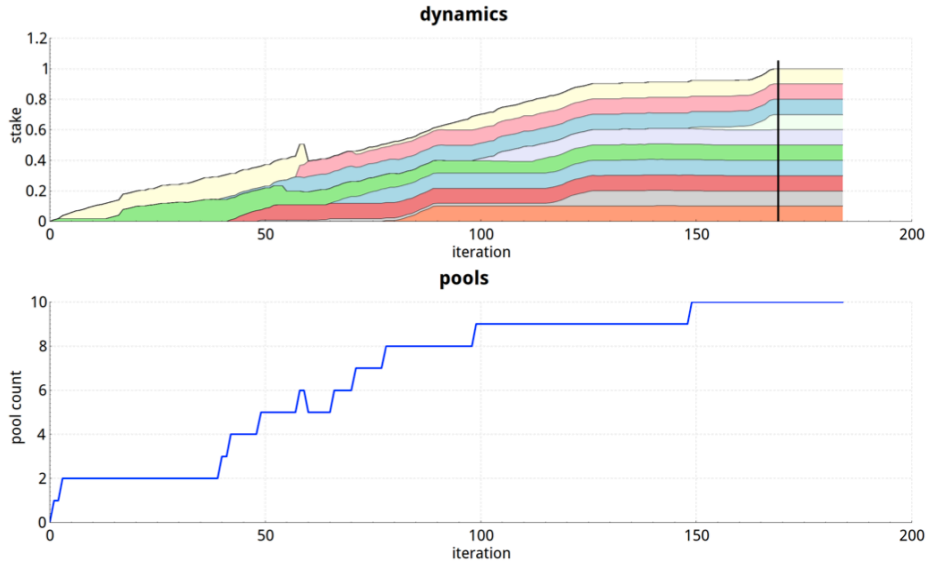

Figure 13: High costs, high stake influence ($c \in [0.05, 0.1]$, $\alpha = 0.5$).

Table 4: High costs, low stake influence ($c \in [0.05, 0.1]$, $\alpha = 0.02$).

| player | rk | crk | srk | cost | margin | player stake | pool stake | reward | desirability |
|---|---|---|---|---|---|---|---|---|---|
| 1 | 4 | 1 | 81 | 0.05010638 | 0.14091746 | 0.00613080 | 0.10000000 | 0.10012262 | 0.042968072279390 |
| 2 | 8 | 2 | 62 | 0.05192430 | 0.10881324 | 0.00693720 | 0.10000000 | 0.10013874 | 0.042968072279376 |
| 3 | 3 | 3 | 39 | 0.05293337 | 0.09055053 | 0.00898080 | 0.10000000 | 0.10017962 | 0.042968072279392 |
| 4 | 10 | 4 | 65 | 0.05373632 | 0.07397050 | 0.00683225 | 0.10000000 | 0.10013665 | 0.042968072279365 |
| 5 | 1 | 5 | 17 | 0.05409406 | 0.06893326 | 0.01216771 | 0.10000000 | 0.10024335 | 0.042968072279396 |
| 6 | 9 | 6 | 26 | 0.05440243 | 0.06209143 | 0.01075376 | 0.10000000 | 0.10021508 | 0.042968072279369 |
| 7 | 7 | 7 | 70 | 0.05441660 | 0.06010553 | 0.00662237 | 0.10000000 | 0.10013245 | 0.042968072279379 |
| 8 | 5 | 8 | 47 | 0.05465632 | 0.05586892 | 0.00835115 | 0.10000000 | 0.10016702 | 0.042968072279383 |
| 9 | 6 | 9 | 96 | 0.05621197 | 0.02127128 | 0.00569447 | 0.10000000 | 0.10011389 | 0.042968072279383 |
| 10 | 2 | 10 | 90 | 0.05651394 | 0.01461751 | 0.00597063 | 0.10000000 | 0.10011941 | 0.042968072279394 |

[11] Decred Contributors. Decred research overview. https://docs.decred.org/research/overview/, 2018.

[12] Phil Daian, Rafael Pass, and Elaine Shi. Snow white: Provably secure proofs of stake. Cryptology ePrint Archive, Report 2016/919, 2016. http://eprint.iacr.org/2016/919.

[13] Bernardo David, Peter Gaži, Aggelos Kiayias, and Alexander Russell. Ouroboros praos: An adaptively-secure, semi-synchronous proof-of-stake protocol. Cryptology ePrint Archive, Report 2017/573, 2017. http://eprint.iacr.org/2017/573. To appear at EUROCRYPT 2018.

[14] John R. Douceur. The sybil attack. In *Revised Papers from the First International Workshop on Peer-to-Peer Systems*, IPTPS '01, pages 251–260, London, UK, UK, 2002. Springer-Verlag.

[15] Ittay Eyal. The miner's dilemma. *CoRR*, abs/1411.7099, 2014.

[16] Ittay Eyal and Emin Gün Sirer. Majority is not enough: Bitcoin mining is vulnerable. In Nicolas Christin and Reihaneh Safavi-Naini, editors, *Financial Cryptography and Data Security - 18th International Conference, FC 2014, Christ Church, Barbados, March 3-7, 2014, Revised Selected Papers*, volume 8437 of *Lecture Notes in Computer Science*, pages 436–454. Springer, 2014.

[17] Giulia Fanti, Leonid Kogan, Sewoong Oh, Kathleen Ruan, Pramod Viswanath, and Gerui Wang. Compounding of wealth in proof-of-stake cryptocurrencies, 2018.

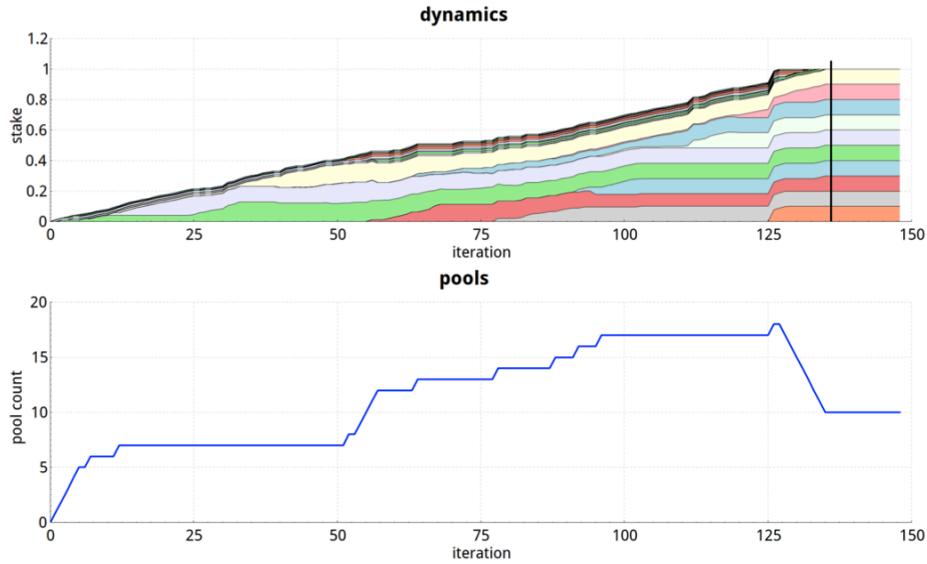

Figure 14: Low costs, low stake influence ($c \in [0.001, 0.002]$, $\alpha = 0.02$), allowing five players to make moves simultaneously, allowing pool moves every 100 rounds.

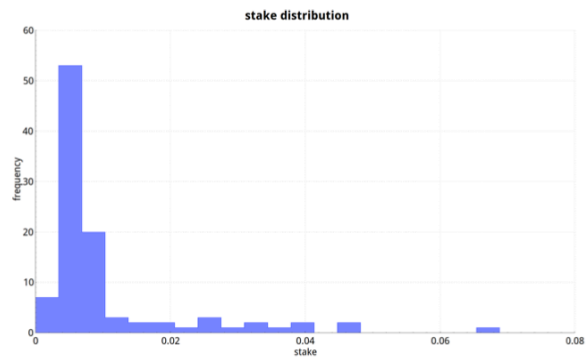

Figure 15: The stake-distribution for parameter 1.1.

[18] C. Fershtman and K. L. Judd. Equilibrium incentives in oligopoly. *American Economic Review*, 77(5):pp. 927 – 940, 1987.

[19] Amos Fiat, Elias Koutsoupias, Katrina Ligett, Yishay Mansour, and Svetlana Olonetsky. Beyond myopic best response (in cournot competition). *Games and Economic Behavior*, 2013.

[20] Ben Fisch, Rafael Pass, and Abhi Shelat. Socially optimal mining pools. In Nikhil R. Devanur and Pinyan Lu, editors, *Web and Internet Economics - 13th International Conference, WINE 2017, Bangalore, India, December 17-20, 2017, Proceedings*, volume 10660 of *Lecture Notes in Computer Science*, pages 205–218. Springer, 2017.

[21] A. Gervais, G. O. Karame, V. Capkun, and S. Capkun. Is bitcoin a decentralized currency? *IEEE Security Privacy*, 12(3):54–60, May 2014.

[22] Ian Grigg. Eos, an introduction. https://eos.io/documents/EOS_An_Introduction.pdf, 2017.

[23] Dimitris Karakostas, Aggelos Kiayias, Christos Nasikas, and Dionysis Zindros. Cryptocurrency egalitarianism: A quantitative approach. Tokenomics, International Conference on Blockchain Economics, Security and Protocols, 2019.

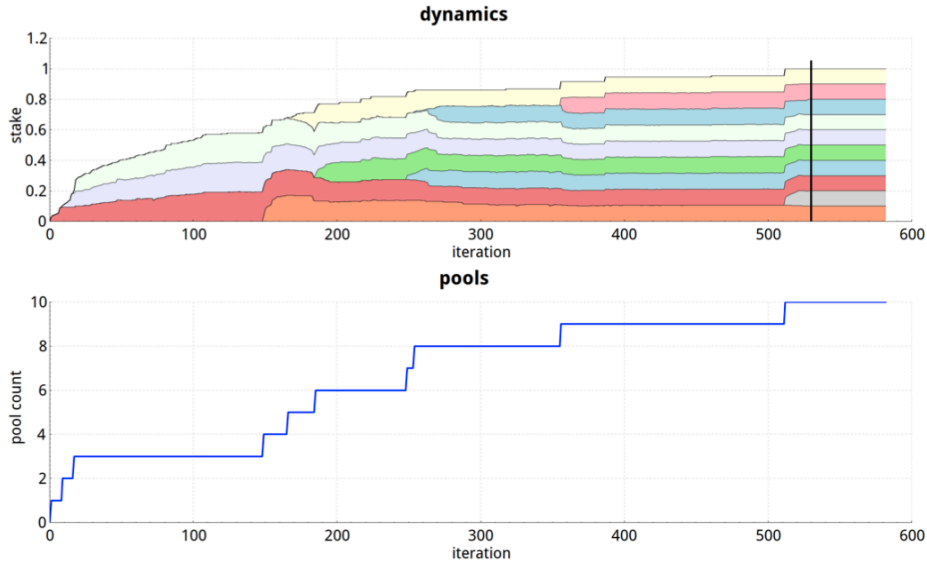

Figure 16: High costs, high stake influence ($c \in [0.001, 0.002]$, $\alpha = 0.5$), Pareto parameter 1.1.

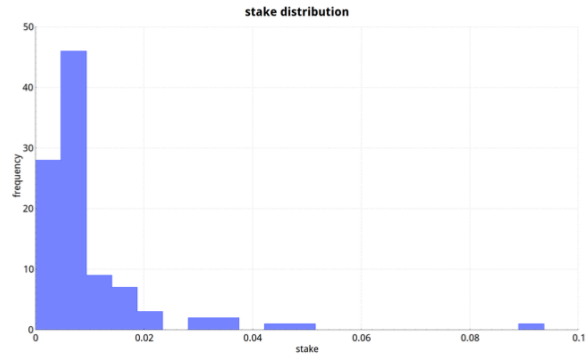

Figure 17: The stake-distribution for parameter 1.3.

[24] Aggelos Kiayias, Elias Koutsoupias, Maria Kyropoulou, and Yiannis Tselekounis. Blockchain mining games. In Vincent Conitzer, Dirk Bergemann, and Yiling Chen, editors, *Proceedings of the 2016 ACM Conference on Economics and Computation, EC '16, Maastricht, The Netherlands, July 24-28, 2016*, pages 365–382. ACM, 2016.

[25] Aggelos Kiayias, Alexander Russell, Bernardo David, and Roman Oliynykov. Ouroboros: A provably secure proof-of-stake blockchain protocol. Cryptology ePrint Archive, Report 2016/889, 2016. http://eprint.iacr.org/2016/889.

[26] Yujin Kwon, Jian Liu, Minjeong Kim, Dawn Song, and Yongdae Kim. Impossibility of full decentralization in permissionless blockchains. *CoRR*, abs/1905.05158, 2019.

[27] Dan Larimer. Delegated proof-of-stake consensus. https://bitshares.org/technology/delegated-proof-of-stake-consensus/, accessed 21.3.2018, 2018.

[28] Nikos Leonardos, Stefanos Leonardos, and Georgios Piliouras. Oceanic games: Centralization risks and incentives in blockchain mining. *CoRR*, abs/1904.02368, 2019.

[29] A. Mas-Colell, J. Green, and M. D. Whinston. *Microeconomic Theory*. Oxford University Press, 1995.

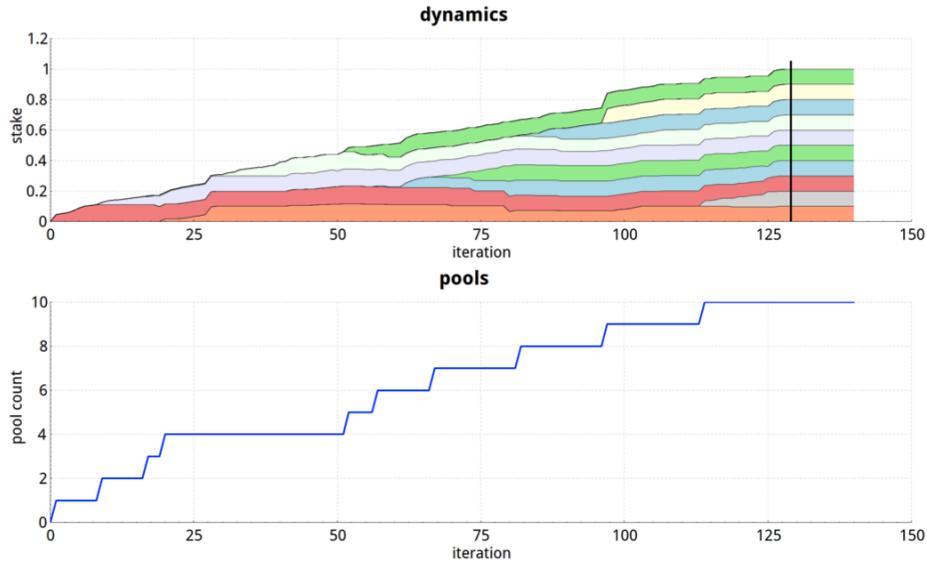

Figure 18: High costs, high stake influence ($c \in [0.001, 0.002]$, $\alpha = 0.5$), Pareto parameter 1.3.

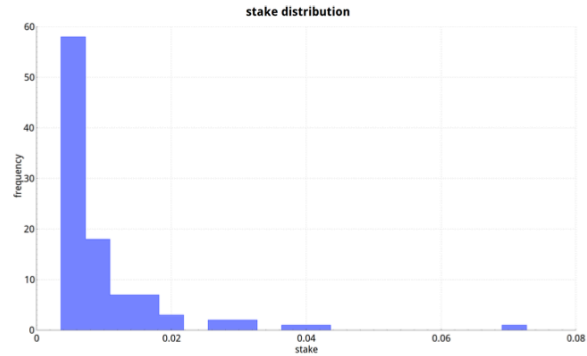

Figure 19: The stake-distribution for parameter 1.5.

[30] Silvio Micali. ALGORAND: The efficient and democratic ledger, 2016.

[31] Noam Nisan, Tim Roughgarden, Eva Tardos, and Vijay V Vazirani. *Algorithmic game theory*. Cambridge University Press, 2007.

[32] A Orda, R Rom, and N Shimkin. Competitive Routing in Multiuse Communication Networks. *IEEE/ACM Transactions on Networking*, 1(5):510–521, 1993.

[33] Guillermo Owen. Game theory academic press. *San Diego*, 1995.

[34] Vilfredo Pareto. *Cours d'Èconomie Politique*. Librairie Droz, Geneva, 1964.

[35] Rafael Pass and Elaine Shi. Fruitchains: A fair blockchain. *IACR Cryptology ePrint Archive*, 2016:916, 2016.

[36] Oran Richman and Nahum Shimkin. Topological Uniqueness of the Nash Equilibrium for Selfish Routing with Atomic Users. *Mathematics of Operations Research*, 32(1):215–232, 2007.

[37] Matteo Romiti, Aljosha Judmayer, Alexei Zamyatin, and Bernhard Haslhofer. A deep dive into bitcoin mining pools: An empirical analysis of mining shares, 2019.

[38] Meni Rosenfeld. Analysis of bitcoin pooled mining reward systems. *CoRR*, abs/1112.4980, 2011.

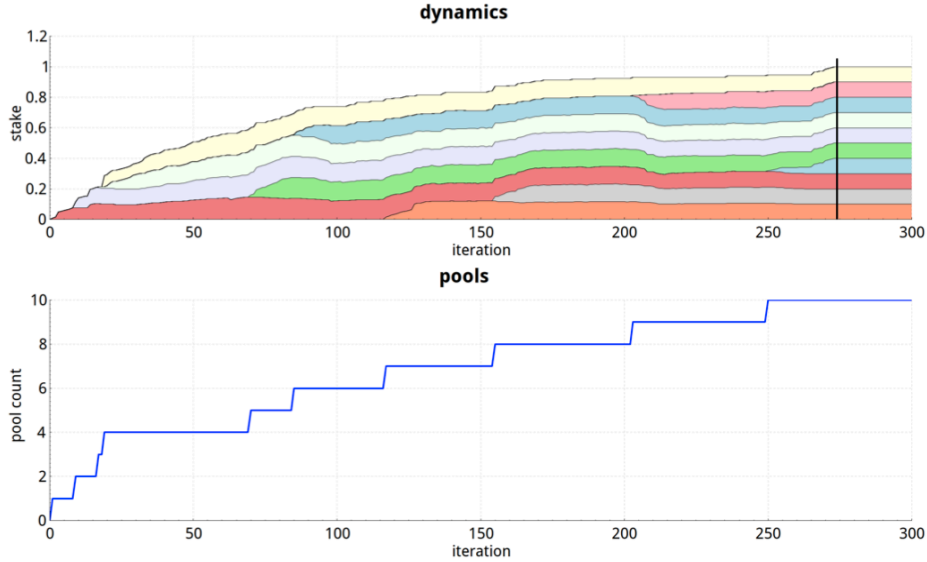

Figure 20: High costs, high stake influence ($c \in [0.001, 0.002]$, $\alpha = 0.5$), Pareto parameter $\alpha = 1.5$.

Table 5: High costs, medium stake influence ($c \in [0.05, 0.1]$, $\alpha = 0.05$).

| player | rk | crk | srk | cost | margin | player stake | pool stake | reward | desirability |
|---|---|---|---|---|---|---|---|---|---|
| 1 | 4 | 1 | 81 | 0.05010638 | 0.14071474 | 0.00613080 | 0.10000000 | 0.10030654 | 0.043136255076695 |
| 2 | 7 | 2 | 62 | 0.05192430 | 0.10917025 | 0.00693720 | 0.10000000 | 0.10034686 | 0.043136255076684 |
| 3 | 6 | 3 | 39 | 0.05293337 | 0.09216778 | 0.00898080 | 0.10000000 | 0.10044904 | 0.043136255076688 |
| 4 | 1 | 4 | 65 | 0.05373632 | 0.07443446 | 0.00683225 | 0.10000000 | 0.10034161 | 0.043136255076709 |
| 5 | 8 | 5 | 17 | 0.05409406 | 0.07262429 | 0.01216771 | 0.10000000 | 0.10060839 | 0.043136255076679 |
| 6 | 9 | 6 | 26 | 0.05440243 | 0.06500457 | 0.01075376 | 0.10000000 | 0.10053769 | 0.043136255076675 |
| 7 | 5 | 7 | 70 | 0.05441660 | 0.06050948 | 0.00662237 | 0.10000000 | 0.10033112 | 0.043136255076690 |
| 8 | 2 | 8 | 47 | 0.05465632 | 0.05736264 | 0.00835115 | 0.10000000 | 0.10041756 | 0.043136255076704 |
| 9 | 10 | 9 | 96 | 0.05621197 | 0.02124899 | 0.00569447 | 0.10000000 | 0.10028472 | 0.043136255076672 |
| 10 | 3 | 10 | 90 | 0.05651394 | 0.01480748 | 0.00597063 | 0.10000000 | 0.10029853 | 0.043136255076696 |

[39] Thomas C Schelling. An essay on bargaining. *The American Economic Review*, 46(3):281–306, 1956.

[40] Thomas C Schelling. *The strategy of conflict*. Harvard university press, 1980.

[41] Okke Schrijvers, Joseph Bonneau, Dan Boneh, and Tim Roughgarden. Incentive compatibility of bitcoin mining pool reward functions. In Jens Grossklags and Bart Preneel, editors, *Financial Cryptography and Data Security*, pages 477–498, Berlin, Heidelberg, 2017. Springer Berlin Heidelberg.

[42] S. D. Sklivas. The strategic choice of managerial incentives. *RAND Journal of Economics*, 18(3):pp. 452 – 458, 1987.

[43] Itay Tsabary and Ittay Eyal. The gap game. *CoRR*, abs/1805.05288, 2018.

[44] J. Vickers. Delegation and the theory of the firm. *Economic Journal*, 95 (Suppl.):pp. 138 – 147, 1985.

[45] William Vickrey. Counterspeculation, auctions, and competitive sealed tenders. *The Journal of finance*, 16(1):8–37, 1961.

Table 6: High costs, high stake influence ($c \in [0.05, 0.1]$, $\alpha = 0.5$).

| player | rk | crk | srk | cost | margin | player stake | pool stake | reward | desirability |
|---|---|---|---|---|---|---|---|---|---|
| 1 | 4 | 1 | 81 | 0.05010638 | 0.08665199 | 0.00613080 | 0.10000000 | 0.10306540 | 0.048370013523060 |
| 2 | 10 | 2 | 62 | 0.05192430 | 0.06158357 | 0.00693720 | 0.10000000 | 0.10346860 | 0.048370013523048 |
| 3 | 2 | 3 | 39 | 0.05293337 | 0.06181530 | 0.00898080 | 0.10000000 | 0.10449040 | 0.048370013523073 |
| 4 | 1 | 5 | 17 | 0.05409406 | 0.06962481 | 0.01216771 | 0.10000000 | 0.10608385 | 0.048370013523074 |
| 5 | 8 | 6 | 26 | 0.05440243 | 0.05109295 | 0.01075376 | 0.10000000 | 0.10537688 | 0.048370013523051 |
| 6 | 6 | 4 | 65 | 0.05373632 | 0.02636475 | 0.00683225 | 0.10000000 | 0.10341613 | 0.048370013523057 |
| 7 | 9 | 8 | 47 | 0.05465632 | 0.02320800 | 0.00835115 | 0.10000000 | 0.10417557 | 0.048370013523048 |
| 8 | 5 | 7 | 70 | 0.05441660 | 0.01072862 | 0.00662237 | 0.10000000 | 0.10331119 | 0.048370013523058 |
| 9 | 3 | 54 | 1 | 0.07842812 | 0.19512770 | 0.07704926 | 0.10000000 | 0.13852463 | 0.048370013523062 |
| 10 | 7 | 19 | 5 | 0.06061469 | 0.02573105 | 0.02052438 | 0.10000000 | 0.11026219 | 0.048370013523056 |

Table 7: High costs, high stake influence ($c \in [0.05, 0.1]$, $\alpha = 0.5$), Pareto parameter 1.1.

| player | rk | crk | srk | cost | margin | player stake | pool stake | reward | desirability |
|---|---|---|---|---|---|---|---|---|---|
| 1 | 1 | 5 | 5 | 0.05140459 | 0.25044417 | 0.03864464 | 0.10000000 | 0.11932232 | 0.050908130626656 |
| 2 | 6 | 19 | 3 | 0.05721165 | 0.22717664 | 0.04616913 | 0.10000000 | 0.12308457 | 0.050908130626627 |
| 3 | 7 | 6 | 9 | 0.05285215 | 0.18214878 | 0.03019669 | 0.10000000 | 0.11509835 | 0.050908130626618 |
| 4 | 2 | 10 | 10 | 0.05366829 | 0.15131031 | 0.02730535 | 0.10000000 | 0.11365268 | 0.050908130626651 |
| 5 | 5 | 7 | 13 | 0.05345880 | 0.11798641 | 0.02235377 | 0.10000000 | 0.11117688 | 0.050908130626630 |
| 6 | 3 | 1 | 26 | 0.05015670 | 0.05887327 | 0.00849889 | 0.10000000 | 0.10424945 | 0.050908130626642 |
| 7 | 8 | 9 | 14 | 0.05365108 | 0.08659790 | 0.01877143 | 0.10000000 | 0.10938572 | 0.050908130626614 |
| 8 | 9 | 3 | 38 | 0.05075232 | 0.03739409 | 0.00727612 | 0.10000000 | 0.10363806 | 0.050908130626611 |
| 9 | 10 | 35 | 2 | 0.06397117 | 0.14410750 | 0.04690152 | 0.10000000 | 0.12345076 | 0.050908130626607 |
| 10 | 4 | 2 | 65 | 0.05017958 | 0.02093887 | 0.00435293 | 0.10000000 | 0.10217647 | 0.050908130626640 |

# Appendices

## A  Proofs

### A.1  Proof for Fair Reward Sharing Scheme

In the proofs we will assume that there exists at most one player with zero cost.

**Proof of Theorem 2**

(I) We will prove it by contradiction. Let suppose that there is an equilibrium where there exist $l$ pools with $l > 1$. We order the pools according to the quantity $\frac{c_i}{\sigma_i}$. Let $\pi_{j_0}$ be the pool with the lowest value of this quantity denoted by $\frac{c_{j_0}}{\sigma_{j_0}}$. Then the members of all the other pools have incentives to delegate their stake to $\pi_{j_0}$. So this cannot be an equilibrium. Specifically let examine a pool member $j$ of a pool $\pi_i$. It holds $\frac{c_i}{\sigma_i} \geq \frac{c_{j_0}}{\sigma_{j_0}}$. The utility of player $j$ if they leave their stake $a_{j,i}$ in $\pi_i$ is $a_{j,i} - a_{j,i} \cdot \frac{c_i}{\sigma_i}$. If they remove their stake from $\pi_i$ and delegates it to $\pi_{j_0}$ their utility will be $a_{j,i} - a_{j,i} \cdot \frac{c_{j_0}}{\sigma_{j_0} + a_{j,i}}$ which is strictly higher. Note that we have assumed that there exists at most one player with zero cost. So if there exists a player with zero cost then this player will be $j_0$.

(II) Firstly we will prove that the joint strategies that satisfy properties (i),(ii) and (iii) as defined by the theorem are indeed equilibria.

The pool leader has no incentives to dissolve their pool because their utility is $s_i - s_i \cdot \frac{c_i}{\sigma_i}$ that is greater or equal to zero, because $1 \geq c_i$ and $\sigma_i = 1$.

The pool members have no incentives to create their own pool because their current utility is $s_j - s_j \cdot \frac{c_i}{\sigma_i}$ and is not lower than the utility they will get if they create their own pool which will be equal to $s_j - c_j$, given that $s_j \cdot c_i \leq c_j$ and $\sigma_i = 1$.

Secondly we will prove that there is no other equilibrium. By Theorem 2 we have that there is no equilibrium with more than one pools. We will prove by contradiction that (a) there is no equilibrium

Table 8: High costs, high stake influence ($c \in [0.05, 0.1]$, $\alpha = 0.5$), Pareto parameter 1.3.

| player | rk | crk | srk | cost | margin | player stake | pool stake | reward | desirability |
|---:|---:|---:|---:|---:|---:|---:|---:|---:|---:|
| 1 | 9 | 4 | 12 | 0.05232917 | 0.11113014 | 0.01640414 | 0.10000000 | 0.10820207 | 0.049663741797333 |
| 2 | 4 | 17 | 5 | 0.05835834 | 0.15566652 | 0.03435679 | 0.10000000 | 0.11717839 | 0.049663741797347 |
| 3 | 6 | 34 | 3 | 0.06297005 | 0.16545389 | 0.04495987 | 0.10000000 | 0.12247994 | 0.049663741797342 |
| 4 | 8 | 2 | 50 | 0.05114911 | 0.03941328 | 0.00570116 | 0.10000000 | 0.10285058 | 0.049663741797336 |
| 5 | 3 | 1 | 93 | 0.05101953 | 0.02301505 | 0.00370642 | 0.10000000 | 0.10185321 | 0.049663741797352 |
| 6 | 5 | 9 | 15 | 0.05490394 | 0.05830496 | 0.01528522 | 0.10000000 | 0.10764261 | 0.049663741797344 |
| 7 | 10 | 3 | 59 | 0.05218122 | 0.01586887 | 0.00529155 | 0.10000000 | 0.10264578 | 0.049663741797329 |
| 8 | 1 | 5 | 47 | 0.05260393 | 0.01749394 | 0.00630391 | 0.10000000 | 0.10315195 | 0.049663741797356 |
| 10 | 7 | 75 | 1 | 0.08299881 | 0.22213113 | 0.09368943 | 0.10000000 | 0.14684472 | 0.049663741797341 |
| 15 | 2 | 35 | 7 | 0.06385628 | 0.01242114 | 0.02828932 | 0.10000000 | 0.11414466 | 0.049663741797354 |

Table 9: High costs, high stake influence ($c \in [0.05, 0.1]$, $\alpha = 0.5$), Pareto parameter 1.5.

| player | rk | crk | srk | cost | margin | player stake | pool stake | reward | desirability |
|---:|---:|---:|---:|---:|---:|---:|---:|---:|---:|
| 1 | 4 | 4 | 12 | 0.05232917 | 0.11860271 | 0.01605699 | 0.10000000 | 0.10802849 | 0.049093236751146 |
| 2 | 2 | 17 | 5 | 0.05835834 | 0.13684114 | 0.03046917 | 0.10000000 | 0.11523459 | 0.049093236751154 |
| 3 | 7 | 2 | 50 | 0.05114911 | 0.05702149 | 0.00642199 | 0.10000000 | 0.10321099 | 0.049093236751120 |
| 4 | 9 | 1 | 93 | 0.05101953 | 0.04099197 | 0.00442242 | 0.10000000 | 0.10221121 | 0.049093236751112 |
| 5 | 8 | 9 | 15 | 0.05490394 | 0.06749872 | 0.01510155 | 0.10000000 | 0.10755078 | 0.049093236751112 |
| 6 | 1 | 34 | 3 | 0.06297005 | 0.12744328 | 0.03846743 | 0.10000000 | 0.11923372 | 0.049093236751156 |
| 7 | 3 | 3 | 59 | 0.05218122 | 0.03413400 | 0.00601885 | 0.10000000 | 0.10300943 | 0.049093236751149 |
| 8 | 5 | 5 | 47 | 0.05260393 | 0.03548387 | 0.00700654 | 0.10000000 | 0.10350327 | 0.049093236751141 |
| 9 | 6 | 6 | 44 | 0.05357812 | 0.01836732 | 0.00717989 | 0.10000000 | 0.10358994 | 0.049093236751129 |
| 30 | 10 | 75 | 1 | 0.08299881 | 0.07968728 | 0.07268579 | 0.10000000 | 0.13634289 | 0.049093236751112 |

with zero pools and (b) there is no equilibrium with one pool but without (i),(ii) and (iii) properties:

(a) Let assume that there is an equilibrium with no pools. Then player $i_0$ for whom it holds $s_{i_0} > c_{i_0}$ can increase their utility by creating their own pool. (b) Let assume that there is an equilibrium with one pool where (i) or (ii) does not hold. If (i) does not hold, then the pool leader of the pool has negative utility and thus they have incentives to dissolve their pool. If (ii) does not hold then a pool member can increase their utility by creating their own pool.

Now we will prove that an arbitrary equilibrium with one pool satisfies (iii). We consider that in an equilibrium player $i_0$ for whom it holds $s_{i_0} > c_{i_0}$ is a pool member or a pool leader of the unique pool $\pi_i$. This holds because if they did not participate at all then they could increase their utility by creating their own pool, but this cannot happen as we are in an equilibrium. As this player is pool leader or pool member the profit of the pool $\pi_i \sigma_i - c_i$ is positive. (This holds because otherwise player $i_0$ has incentives to create their own pool.) So other players cannot have chosen to not participate at all in the equilibrium (which means that (iii) holds) because otherwise they could increase their utility by delegating to $\pi_i$.

**Theorem 7.** *If it holds $c_i > 1$ for all players then there is exactly one equilibrium where no pool has been created.*

*Proof.* It can be trivially proved that indeed there is an equilibrium where no pool exists, because if a player $i$ decides to create a pool then their utility will become negative ($a_{i,i} - c_i$) that is strictly lower compared to their utility when they do not participate at all that is zero. In addition we can easily prove by contradiction that there is no equilibrium with one or more pools. Specifically let suppose that there exists. Then the pool leader of the pool has negative utility, so they have incentives to dissolve their pool. □

## A.2 Proofs when $(r(\sigma, \lambda) - c)/\sigma$ is strictly increasing or decreasing

**Proof of Theorem 3**

I) We will prove it by contradiction. Let assume that there is an equilibrium where there are $l > 1$ pools. Then we order the pools according to the quantity $\frac{r(\sigma_i, a_{i,i}) - c_i}{\sigma_i}$. Let $\pi_{j_0}$ be the pool with the highest value. Then this cannot be an equilibrium given that the members of the other pools have incentives to delegate their stake to $\pi_{j_0}$. In more detail, the utility of a player $j$ that leaves their stake to pool $\pi_i$ is $\frac{r(\sigma_i, a_{i,i}) - c_i}{\sigma_i} \cdot a_{j,i}$ and if they remove it and reallocate it to pool $\pi_{j_0}$ then it becomes $\frac{r(\sigma_{j_0} + a_{j,i}, a_{j_0,j_0}) - c_{j_0}}{\sigma_{j_0} + a_{j,i}} \cdot a_{j,i}$ that is strictly higher.

II) Let $f \in (1, \infty)$. We will find an assignment of costs and stake to the players so that there is no equilibrium with a number of pools smaller than the number of players $n$. In more detail: (i) the assignment of stake will be such that each player has stake $\frac{1}{n}$ (ii) the assignment of costs will be arbitrary except the minimum cost $c_{min}$ and the maximum cost $c_{max}$ that we will determine. Specifically we take $c_{max}$ such that $max\{r(\frac{s_{min}}{f}) - \frac{1}{f} \cdot r(s_{min}), 0\} < c_{max} < r(\frac{s_{min}}{f})$ and we will determine $c_{min}$ as follows:

- It holds $\frac{r(\frac{s_{min}}{f}) - c_{max}}{\frac{s_{min}}{f}} > \frac{r(\frac{1}{n}) - c_{max}}{\frac{1}{n}}$ because $r(\frac{s_{min}}{f}) > c_{max}$, $\frac{s_{min}}{f} < \frac{1}{n}$ and $\frac{r(\sigma, \lambda) - c}{\sigma}$ as a function of $\sigma$ is strictly decreasing.

- Let $r_0$ such that $y = \frac{r(\frac{s_{min}}{f}) - c_{max}}{\frac{s_{min}}{f}} > r_0 > \frac{r(\frac{1}{n}) - c_{max}}{\frac{1}{n}}$.

- We define the function $g(x) = \frac{r(\frac{1}{n}) - x}{\frac{1}{n}}$ which is continuous in $[c, c_{max}]$, where $c$ such that $\frac{r(\frac{s_{min}}{f}) - c_{max}}{\frac{s_{min}}{f}} = \frac{r(\frac{1}{n}) - c}{\frac{1}{n}}$.

- We know that there exists $c \in (0, c_{max})$ such that $y = g(c)$ by the *intermediate value theorem* because (i) $g(c_{max}) < y < g(0)$ given that $max\{r(\frac{s_{min}}{f}) - \frac{1}{f} \cdot r(s_{min}), 0\} < c_{max} < r(\frac{s_{min}}{f})$ and (ii) $g(x)$ continuous in $[0, c_{max}]$. Note that $s_{min} = \frac{1}{n}$ given that all players have stake $\frac{1}{n}$.

- By the *intermediate value theorem* again there is $x \in (c, c_{max})$ such that $g(c) > g(x) = r_0 > g(c_{max})$.

- We set $c_{min} = x$ the minimum cost among all players. Note that it holds $g(c) = \frac{r(\frac{s_{min}}{f}) - c_{max}}{\frac{s_{min}}{f}} > \frac{r(\frac{1}{n}) - c_{min}}{\frac{1}{n}} = g(c_{min})$.

Now we will prove that for these values of $c_{min}, c_{max}$ and assignment of stake ($1/n$ for each player) there is no equilibrium with fewer than $n$ pools.

- Firstly we will prove $C \Rightarrow A \cup B$ where $C$ the event where there exist fewer than $n$ pools, $A$ the event where there exists at least one player that has left some of their stake unallocated and $B$ the event where there exists at least one pool with stake more than $1/n$. In order to prove $C \Rightarrow A \cup B$ we will prove $\neg(A \cup B) \Rightarrow \neg C$ which is equivalent to $\neg A \cap \neg B \Rightarrow \neg C$. Let assume that there is no stake unallocated (so the total stake that is delegated to pools is 1) and all pools have stake at most $1/n$. If $x$ is the number of the pools then the total stake that is delegated is at most $x \cdot \frac{1}{n}$ which means that $x$ is at least $n$.

- Now in order to prove that there is no equilibrium with fewer than $n$ pools it is sufficient to prove that there is no equilibrium where the event $A$ or the event $B$ happens.

- Firstly we will prove by contradiction that there is no equilibrium where the event $B$ happens. Let assume that there exists such an equilibrium where a pool $\pi_i$ has stake more than $1/n$. Then this pool has pool members given that $s_i = \frac{1}{n}$. Then player $j$ who has delegated to this pool stake $a_{j,i} \geq \frac{s_{min}}{f}$ has incentives to remove stake $\frac{s_{min}}{f}$ from pool $\pi_i$ and create their own pool. This happens because the utility of player $j$ from pool $\pi_i$ for this part of their stake is smaller than $\frac{r(\frac{1}{n}) - c_{min}}{\frac{1}{n}} \cdot \frac{s_{min}}{f}$ which is smaller than $\frac{r(\frac{s_{min}}{f}) - c_{max}}{\frac{s_{min}}{f}} \cdot \frac{s_{min}}{f}$ and thus smaller than their utility if they create their own pool with stake $\frac{s_{min}}{f}$.

- We can easily prove by contradiction that there is no equilibrium where the event $A$ happens. Let assume that there is a player $j$ that has left some of their stake unallocated. If this part of stake is $\frac{s_{min}}{f}$ or more then they can increase their utility by creating their own pool given that $r(\frac{s_{min}}{f}) - c_{max} > 0$. If this player has some stake $x$ smaller than $\frac{s_{min}}{f}$ unallocated then this means that either (i) they have allocated stake to another pool and thus there exists a pool with stake more than $1/n$ (this leads to contradiction as we have proved above) or (ii) they have created their own pool. In the latter case this player has incentives to include the unallocated part of stake to this pool given that the function $r$ is strictly increasing ($r(s_j) > r(s_j - x)$). Note that this pool will not have other members as we have proved above.

### A.3 Proof of Theorem 5

*Proof.* Consider the the equilibrium of Theorem 4, where the $k$ players with the highest potential profit $P(s_i, c_i)$ have a saturated pool, cf. Proposition 2.

Let $n$ the number of players in the stake-pool game, some of which in a subset $A$ are Sybil players controlled by an agent with stake $\tilde{s}$ and cost $\tilde{c}$. For simplicity we drop the superscript $A$ from $s_i^A, c_i^A$.

Let us suppose first that the agent creates $t > 1$ Sybil players with stake $s = \tilde{s}/t$ each, and claims cost $c$ for each one equal to (i) $\frac{c_{\text{fake}}}{k/2}$, in the first case (where $t = k/2$) where $c_{\text{fake}}$ is some arbitrary cost potentially below $\tilde{c}$, and (ii) $\tilde{c}/t$, in the second case. The objective of the agent in the first case is to create $k/2$ saturated pools so that it musters the highest possible influence in the system, and in the second case to maximize its utility by sharing the same server for all its Sybil players. We provide a lower bound in both cases for the stake the agent needs in order to create $t$ saturated pools. After that, we generalize the above result by allowing the attacker to split its stake and cost arbitrarily among the Sybil players. Whenever needed, without loss of generality, we will break any ties in favor of the adversary. Firstly we will prove a lemma regarding when the attacker succeeds in creating $t$ pools when the Sybil players are identical and play rationally without colluding.

**Lemma 3.** *An attacker with $t$ identical Sybil players, with stake $s$ and cost $c$ each one, controls $t$ saturated pools at the Nash equilibrium with rank at most $k$ if and only if $P(s, c) \geq P(s_{k-t+1}, c_{k-t+1})$*

*Proof.*
- ($\Leftarrow$) If $P(s, c) \geq P(s_{k-t+1}, c_{k-t+1})$ then at most $k - t$ players may have higher $P(s_i, c_i)$ value than the Sybil players and thus all the Sybil players will have a saturated pool.

- ($\Rightarrow$) if $P(s, c) < P(s_{k-t+1}, c_{k-t+1})$ then there would exist at least $k - t + 1$ players with $P(s_i, c_i)$ higher than the agent's players hence superseding all the players in the pool rankings. Thus at most $t - 1$ Sybil players can have a saturated pool.

This completes the proof of the lemma. □

Now we observe that

$$(11) \quad P(s, c) \geq P(s_{k-t+1}, c_{k-t+1}) \iff s \geq s_{k-t+1} - \frac{1}{R} \cdot (c_{k-t+1} - c) \cdot (1 + \frac{1}{\alpha})$$

Finally, with respect to the two scenarios, we have the following. In the first scenario, the attacker does not care about cost and hence can set $c = 0$.

$\tilde{s} < \frac{k}{2} \cdot (s_{k/2+1} - \frac{c_{\max}}{R} \cdot (1 + \frac{1}{\alpha}))$, we obtain

$$s = \tilde{s}/t < s_{k/2+1} - \frac{c_{\max}}{R} \cdot (1 + \frac{1}{\alpha}) \le s_{k-t+1} - \frac{1}{R} \cdot (c_{k-t+1} - c) \cdot (1 + \frac{1}{\alpha})$$

by setting $t = k/2$ and observing that $c_{k-t+1} - c \le c_{\max}$, from which we obtain that $P(s,c) < P(s_{k/2+1}, c_{k/2+1})$ and hence the adversary fails to control $k/2$ pools in the equilibrium.

In the second scenario where it holds that $c = \tilde{c}/t$, and $\tilde{s} < t \cdot \left(s_{k-t+1} - \frac{(c_{\max} - \tilde{c}/t)}{R} \cdot (1 + \frac{1}{\alpha})\right)$ we obtain

$$s = \tilde{s}/t < s_{k-t+1} - \frac{(c_{\max} - \tilde{c}/t)}{R} \cdot (1 + \frac{1}{\alpha}) \le s_{k-t+1} - \frac{1}{R} \cdot (c_{k-t+1} - c) \cdot (1 + \frac{1}{\alpha})$$

recalling that $c = \tilde{c}/t$ and $c_{\max} \ge c_{k-t+1}$. It follows that $P(s,c) < P(s_{k-t+1}, c_{k-t+1})$ and hence the adversary fails to control $t$ pools in the equilibrium.

In order to generalize the above results so that they hold even if the stake and the cost of the attacker is split arbitrarily among the Sybil players we will prove by contradiction the following lemma.

**Lemma 4.** *If an agent splits its stake and cost among the Sybil players so that it succeeds in creating $t$ saturated pools at the Nash equilibrium then it would succeed even if the split of stake and cost is done equally among the Sybil players.*

*Proof.* We will prove the theorem by contradiction. We will assume that the attacker splits its stake and cost among the Sybil players so that it has $t$ saturated pools at the Nash equilibrium but if it splits them equally then it won't. Let $\tilde{s}_1, ..., \tilde{s}_t$ and $\tilde{c}_1, ..., \tilde{c}_t$ the stake and the cost of the Sybil players. It holds $\tilde{s}_1 + ... + \tilde{s}_t = \tilde{s}$ and $\tilde{c}_1 + ... + \tilde{c}_t \le \tilde{c}$ ("=" in the case of a utility maximizing agent).

If the attacker has $t$ saturated pools at the Nash equilibrium, then player $k - t + 1$ will not have a pool. If the $(k - t + 1)$-th player had a pool, then also players $1, 2..., k - t$ would have pools and the attacker would have at most $t - 1$ pools. Recall that we break ties in favor of the attacking agent. As a result $P(s_{k-t+1}, c_{k-t+1})$ is smaller or equal than the potential profit of all the Sybil players (recall that all the Sybil players have a pool and that at the Nash equilibrium the $k$ players with the highest potential profit have a pool), otherwise $k - t + 1$ would have a pool. Thus it holds by equation 11 that $\forall i : \tilde{s}_i \ge s_{k-t+1} - \frac{1}{R} \cdot (c_{k-t+1} - \tilde{c}_i) \cdot (1 + \frac{1}{\alpha})$. Summing for all $i$, we obtain that $\tilde{s} \ge tX + \frac{1}{R}(1 + \frac{1}{\alpha}) \sum_{i=1}^{t} \tilde{c}_i$, where $X = s_{k-t+1} - \frac{1}{R} \cdot c_{k-t+1} \cdot (1 + \frac{1}{\alpha})$.

If, on the other hand, the agent splits its stake equally and declares the same cost $c'$, for some $c' \ge 0$, for all Sybil players but it does not succeed in having $t$ saturated pools then by Lemma 3 and equation 11 we have $\tilde{s}/t < s_{k-t+1} - \frac{1}{R} \cdot (c_{k-t+1} - c') \cdot (1 + \frac{1}{\alpha})$, which implies $\tilde{s} < t \cdot X + t \cdot \frac{1}{R}(1 + \frac{1}{\alpha})c'$. Combining the above constraints on $\tilde{s}$, we obtain that $\sum_{i=1}^{t} \tilde{c}_i < t \cdot c'$.

In the case of a utility maximizer, we have that $\sum_{i=1}^{t} \tilde{c}_i = \tilde{c}$ and since we can set $c' = \tilde{c}/t$, we obtain a contradiction. In the case of a non-utility maximizer, on the other hand, we can set $c' = \sum_{i=1}^{t} \tilde{c}_i / t$, hence also obtaining a contradiction. □

Combining the above results, we arrive at the proof of the theorem.

□

## A.4 Proof of Theorem 6

$$Pr(S_1 > \frac{k}{2} \cdot S_{\frac{k}{2}+1}) \approx \int_{-\infty}^{+\infty} Pr(A/S_{\frac{k}{2}+1} = t) \cdot f_{S_{\frac{k}{2}+1}}(t) dt, \text{ by the Continuous Law of Alternatives (see Section 6.4 in [2])}$$

$$= \int_{\theta}^{\frac{2T}{k}} Pr(A/S_{\frac{k}{2}+1} = t) \cdot F'_{S_{\frac{k}{2}+1}}(t) dt$$

$$= \int_{\theta}^{\frac{2T}{k}} Pr(S_1 > \frac{k}{2} \cdot t) \cdot F'_{S_{\frac{k}{2}+1}}(t) dt$$

$$= \int_{\theta}^{\frac{2T}{k}} (1 - F_{S_1}(\frac{k}{2} \cdot t)) \cdot F'_{S_{\frac{k}{2}+1}}(t) dt$$

$$= \int_{\theta}^{\frac{2T}{k}} (1 - F_X^n(\frac{k}{2} \cdot t)) \cdot F'_{S_{\frac{k}{2}+1}}(t) dt$$

$$\leq \int_{\theta}^{\frac{2T}{k}} F'_{S_{\frac{k}{2}+1}}(t) dt$$

$$= F_{S_{\frac{k}{2}+1}}(\frac{2T}{k}) - F_{S_{\frac{k}{2}+1}}(\theta)$$

$$= F_{S_{\frac{k}{2}+1}}(\frac{2T}{k})$$

$$= F_{X_{n-\frac{k}{2}}}(\frac{2T}{k})$$

$$= \sum_{j=n-\frac{k}{2}}^{n} [\binom{n}{j} \cdot F_X^j(\frac{2T}{k}) \cdot \left(1 - F_X(\frac{2T}{k})\right)^{n-j}]$$

$$= Pr(X_B \geq n - \frac{k}{2})$$

$$= Pr(X_B \geq (1+\delta) \cdot \mu)$$

$$\leq e^{-\delta^2 \mu / 3}$$

where $\delta = \left(\frac{1 - (\frac{\theta}{T})^\alpha}{1 - (\frac{\theta \cdot k}{2 \cdot T})^\alpha}\right) \cdot (1 - \frac{k}{2n}) - 1 > 0$ and $\mu = n \cdot F_X(\frac{2T}{k})$.

## A.5 Analysis of the Two-stage Game

Now we will describe a set of joint strategies that (i) are approximate non-myopic Nash equilibria of the outer game and (ii) have the characteristic that in the inner games defined by these joint strategies, all the equilibria form $k$ saturated pools. Recall that a pool is *saturated* when its stake is at least $\beta$. The pool leaders of these pools in these equilibria of the inner game are again the players with the highest values $P(s_i, c_i)$. Note that if all players activated a pool of size $1/k$ with the same margin and their whole stake, then the $k$ pools with the highest potential profit ($P(s_i, c_i)$) would give the highest utility to their members. The intuition for how the set of margins of these joint strategies is determined is the following: The $k$ players with the highest values $P(s_i, c_i)$ set the maximum margin they can, so that their pools belong to the $k$ most desirable pools (the pools with the highest desirability), no matter which margins the other players set. Note that in this model, these players want their pools to have strictly higher desirability than the potential pool of player $k+1$, because in a tie in ranking, they might otherwise lose. In more detail: Let $G = \{1, \ldots, k\}$ be the set of those $k$ players with the highest $P(s_i, c_i)$, $\epsilon = P(s_k, c_k) - P(s_{k+1}, c_{k+1})$ and $\epsilon_1 = P(s_{k+1}, c_{k+1}) - P(s_{k+2}, c_{k+2})$. By assumption $\epsilon, \epsilon_1 > 0$.

For $\epsilon'$ such that $0 < \epsilon' < \min\{\epsilon, P(s_{k+1}, c_{k+1})\}$ and $\alpha$ such that $\frac{s_{k+1}}{\beta} < \alpha < 1$, we define $\vec{m}^*(\epsilon', \alpha)$ as

follows:

$$m_i^*(\epsilon', \alpha) = \begin{cases} \frac{P(s_i,c_i) - P(s_{k+1},c_{k+1}) - \epsilon' \cdot (1-\alpha)}{P(s_i,c_i)} & \text{when } i \in G, \\ \frac{\epsilon' \cdot \alpha}{P(s_{k+1},c_{k+1})} & \text{when } i = k+1, \\ 0 & \text{elsewhere,} \end{cases}$$

and let $\vec{\lambda}^*$ be the vectors with $\lambda_i^* = s_i$ for $i \in [n]$.

Note that margins are well defined because $0 \leq m_i^* < 1$. We prove that: (i) For each $\epsilon'$ such that $0 < \epsilon' < \min\{\epsilon, P(s_{k+1}, c_{k+1}), \epsilon_1\}$, the joint strategies $(\vec{m}^*(\epsilon', \alpha), \vec{\lambda}^*)_{\frac{s_{k+1}}{\beta} < \alpha < 1}$ are $\epsilon'$-non-myopic Nash equilibria of the outer game. So we prove that for every $\epsilon'$ as defined above, there is a class of joint strategies that are $\epsilon'$-non-myopic Nash equilibria of the outer game (Theorem 9). (ii) For each $\epsilon'$ such that $0 < \epsilon' < \min\{\epsilon, P(s_{k+1}, c_{k+1}), \epsilon_1\}$ and $\alpha$ such that $\frac{s_{k+1}}{\beta} < \alpha < 1$, all the equilibria of the inner game determined by joint strategy $(\vec{m}^*(\epsilon', \alpha), \vec{\lambda}^*)$ form $k$ saturated pools (Theorem 8). (iii) In the general case (for any $\alpha$), the pool leaders of the $k$ saturated pools described above are the players of $G$, which are the players with the highest $P(s, c)$. If $\alpha = 0$, the players with highest $P(s, c)$ are the players with lowest cost, so this achieves an optimum in terms of social welfare in this case, since it minimizes the costs of running the system (Theorem 8).

First, we will state a basic lemma that we will use in the proofs of the following lemmas and theorems. This lemma is a generalization of Lemma 2 and intuitively says that according to any joint strategy of any inner game, the utility that a player takes from allocating stake $s$ to other active pools is upper bounded by $D_{\max} \cdot s/\beta$, where $D_{\max}$ is the maximum desirability of all the other active pools. Formally:

**Lemma 5.** *For every joint strategy of the outer game $(\vec{m}, \vec{\lambda})$ and for every joint strategy $\vec{S}^{(\vec{m}, \vec{\lambda})}$ of the inner game determined by $(\vec{m}, \vec{\lambda})$, it holds: For every player $j$ that has allocated stake $s$ to other active pools $\sum_{i \in [n] \smallsetminus \{j\}: i \text{ active}} u_{j,i}(\vec{S}^{(\vec{m}, \vec{\lambda})}) \leq D_{\max} \cdot \frac{s}{\beta}$, where $D_{\max}$ is the maximum desirability according to $\vec{S}^{(\vec{m}^*, \vec{\lambda}^*)}$ of all the other active pools.*

Its proof is similar to the proof of Lemma 2. Recall that inactive pools in this model have desirability zero and their members (if they exist) take utility zero from these pools.

### A.5.1 Equilibria of the Inner Game

**Theorem 8.** *For every $\epsilon'$: $0 < \epsilon' < \min\{\epsilon, P(s_{k+1}, c_{k+1}), \epsilon_1\}$ and for every $\alpha$: $\frac{s_{k+1}}{\beta} < \alpha < 1$, it holds: A joint strategy $\vec{S}^{(\vec{m}^*(\epsilon', \alpha), \vec{\lambda}^*)}$ of the inner game determined by $(\vec{m}^*(\epsilon', \alpha), \vec{\lambda}^*)$ is a Nash equilibrium if and only if it forms $k$ active, saturated pools, whose pool leaders belong to $G$.*

*Proof.* This can be proved by the following two Lemmas 6, 7. □

**Lemma 6.** *For every $\epsilon'$: $0 < \epsilon' < \min\{\epsilon, P(s_{k+1}, c_{k+1}), \epsilon_1\}$ and for every $\alpha$: $\frac{s_{k+1}}{\beta} < \alpha < 1$, it holds: In an inner game determined by $(\vec{m}^*(\epsilon', \alpha), \vec{\lambda}^*)$, joint strategies $\vec{S}^{(\vec{m}^*(\epsilon', \alpha), \vec{\lambda}^*)}$ that form $k$ active saturated pools, whose pool leaders belong to $G$, are Nash equilibria.*

*Proof.* We omit this proof because it is similar to the proof of Theorem 4. □

**Lemma 7.** *For every $\epsilon'$: $0 < \epsilon' < \min\{\epsilon, P(s_{k+1}, c_{k+1}), \epsilon_1\}$ and for every $\alpha$: $\frac{s_{k+1}}{\beta} < \alpha < 1$ it holds: In an inner game determined by $(\vec{m}^*(\epsilon', \alpha), \vec{\lambda}^*)$, joint strategies $\vec{S}^{(\vec{m}^*(\epsilon', \alpha), \vec{\lambda}^*)}$ that do not form $k$ active saturated pools, whose pool leaders belong to $G$, are not a Nash equilibrium.*

*Proof Sketch.* Let $\epsilon'$: $0 < \epsilon' < \min\{\epsilon, P(s_{k+1}, c_{k+1}), \epsilon_1\}$ and $\alpha$: $\frac{s_{k+1}}{\beta} < \alpha < 1$. For simplicity we write $\vec{m}^*$ instead of $\vec{m}^*(\epsilon', \alpha)$. First, we prove that there is no Nash equilibrium joint strategy $S^{(\vec{m}^*, \vec{\lambda}^*)}$ for which there exist one or more players in $G$ that have not chosen to activate their own pools. Second, we prove that there no Nash equilibrium joint strategy $S^{(\vec{m}^*,\vec{\lambda}^*)}$ for which there exist one or more players $\notin G$ that have activated their pools. Last, we prove that there no Nash equilibrium joint strategy $S^{(\vec{m}^*,\vec{\lambda}^*)}$ for which there exist one or more players who have allocated some of their stake to a pool with total stake more than $\beta$ or for which there exists some stake that is unallocated or has been allocated to an inactive pool. □

### A.5.2 Equilibria of the Outer Game

**Theorem 9.** *For every $\epsilon'$: $0 < \epsilon' < \min\{\epsilon, P(s_{k+1}, c_{k+1}), \epsilon_1\}$ and for every $\alpha$: $\frac{s_{k+1}}{\beta} < \alpha < 1$ it holds: Joint strategy $(\vec{m}^*(\epsilon', \alpha), \vec{\lambda}^*)$ is an $\epsilon'$-non-myopic Nash equilibrium of the outer game.*

*Proof.* Let $\epsilon'$: $0 < \epsilon' < \min\{\epsilon, P(s_{k+1}, c_{k+1}), \epsilon_1\}$ and $\alpha$: $\frac{s_{k+1}}{\beta} < \alpha < 1$. For simplicity we write $\vec{m}^*$ instead of $\vec{m}^*(\epsilon', \alpha)$.

We will prove that $\forall i \in [n]$ and $\forall (m_i, \lambda_i) \neq (m_i^*, \lambda_i^*)$, it holds that

$$u_i^{\text{outer,up}}(m_i, \vec{m}_{-i}^*, \lambda_i, \vec{\lambda}_{-i}^*) \leq u_i^{\text{outer,low}}(\vec{m}^*, \vec{\lambda}^*) + \epsilon'.$$

Specifically, we will examine the following cases for $i \in G$:

- $(m_i, \lambda_i)$, where $1 \geq m_i > m_i^*$ and $0 < \lambda_i \leq s_i = \lambda_i^*$.
- $(m_i, \lambda_i)$, where $0 \leq m_i < m_i^*$ and $0 < \lambda_i \leq s_i = \lambda_i^*$.
- $(m_i, \lambda_i)$, where $m_i = m_i^*$ and $0 < \lambda_i < s_i = \lambda_i^*$.
- $(m_i, \lambda_i) = (m_i, 0)$.

For $i \notin G$ we will examine the case $(m_i, \lambda_i) \neq (m_i^*, \lambda_i^*)$.

These are all the ways in which players can change the strategy described in the theorem. For each case we will prove that in the inner game that is determined by $(m_i, \vec{m}_{-i}^*, \lambda_i, \vec{\lambda}_{-i}^*)$, there is no equilibrium in which the utility of $i$ is higher than $u_i^{\text{outer,low}}(\vec{m}^*, \vec{\lambda}^*) + \epsilon'$.

The cases are described below:

- First we will prove that no player $i \in G$ has incentives to increase their margin more than $m_i^*$ in the outer game, regardless of $\lambda_i > 0$. By Lemmas 6, 7 the only equilibria of the inner game determined by $(\vec{m}*, \vec{\lambda}*)$ are that ones where $k$ saturated, active pools have been activated by players in $G$. So for every $i \in G$ we have:

$$\begin{aligned}
&u_i^{\text{outer,low}}(\vec{m}^*, \vec{\lambda}^*) \\
&= u_i^{\text{outer,up}}(\vec{m}^*, \vec{\lambda}^*) \\
&= (m_i^* + (1 - m_i^*) \cdot \frac{s_i}{\beta}) \cdot P(s_i, c_i) \\
&> \frac{s_i}{\beta} \cdot (1 - m_i^*) \cdot P(s_i, c_i) \\
&\stackrel{i \in G}{=} \frac{s_i - \lambda_i}{\beta} \cdot (P(s_{k+1}, c_{k+1}) + \epsilon' \cdot (1 - \alpha)) \\
&+ \frac{\lambda_i}{\beta} \cdot (P(s_{k+1}, c_{k+1}) + \epsilon' \cdot (1 - \alpha)),
\end{aligned}$$

where $0 < \lambda_i \leq s_i$.

1. If an $i \in G$ increases their margin by choosing $m_i$ with $\frac{P(s_i,c_i)-P(s_{k+1},c_{k+1})+\epsilon'\cdot\alpha}{P(s_i,c_i)} \geq m_i > m_i^*$ and chooses arbitrary $\lambda_i$ with $0 < \lambda_i \leq s_i$, then:

$$u_i^{\text{outer,up}}(m_i, \vec{m}_{-i}^*, \lambda_i, \vec{\lambda}_{-i}^*)$$
$$\leq \max\{(m_i + (1-m_i)\cdot\frac{\lambda_i}{\beta})\cdot P(s_i,c_i)$$
$$+ \frac{s_i - \lambda_i}{\beta}\cdot(P(s_{k+1},c_{k+1})$$
$$+ \epsilon'\cdot(1-\alpha)), \frac{s_i}{\beta}\cdot(P(s_{k+1},c_{k+1}) + \epsilon'\cdot(1-\alpha))\}$$
$$\leq m_i^*\cdot P(s_i,c_i) + \epsilon' + (1-m_i^*)\cdot\frac{s_i}{\beta}\cdot P(s_i,c_i)$$
$$\leq u_i^{\text{outer,low}}(\vec{m}^*, \vec{\lambda}^*) + \epsilon'.$$

   Note that in the best case there is an equilibrium where (i) $i$ has not activated their own pool, and their utility is at most $\frac{s_i}{\beta}\cdot(P(s_{k+1},c_{k+1})+\epsilon'\cdot(1-\alpha))$ by Lemma 5 or (ii) $i$ has activated their own pool, that belongs to the $k$ most desirable pools, and has allocated the remaining stake to the active pools with the highest desirability, which are the pools of players in $G$. The utility that these pools will give them will be at most $\frac{s_i-\lambda_i}{\beta}\cdot(P(s_{k+1},c_{k+1})+\epsilon'\cdot(1-\alpha))$ by Lemma 5.
   
   If there is no equilibrium, recall that
   
   $$u_i^{\text{outer,up}}(m_i, \vec{m}_{-i}^*, \lambda_i, \vec{\lambda}_{-i}^*) = -\infty$$
   .

2. If an $i \in G$ increases their margin by choosing $m_i$ with $1 \geq m_i > \frac{P(s_i,c_i)-P(s_{k+1},c_{k+1})+\epsilon'\cdot\alpha}{P(s_i,c_i)}$ and chooses arbitrary $\lambda_i > 0$, then we can prove that there is no equilibrium in the inner game determined by $(m_i, \vec{m}_{-i}^*, \lambda_i, \vec{\lambda}_{-i}^*)$, where the non-myopic utility of $i$ will be higher than their current lower utility of the outer game denoted by $u_i^{\text{outer,low}}(\vec{m}^*, \vec{\lambda}^*)$.

   This happens because in the inner game determined by $(m_i, \vec{m}_{-i}^*, \lambda_i, \vec{\lambda}_{-i}^*)$ we can prove that there is no an equilibrium where player $k+1$ and the other players of $G$ have not activated their own pools (note that the desirability of $\pi_{k+1}$ and of the pools whose pool leaders belong to $G$, when they are active, are strictly higher than the desirability of $\pi_i$, because $m_i > \frac{P(s_i,c_i)-P(s_{k+1},c_{k+1})+\epsilon'\cdot\alpha}{P(s_i,c_i)}$).

   So in the best case, in the inner game determined by $(m_i, \vec{m}_{-i}^*, \lambda_i, \vec{\lambda}_{-i}^*)$ (i) there is an equilibrium where player $i$ has activated their own pool with rank worse than $k$ and has delegated their remaining stake $(s_i - \lambda_i)$ to pools whose pool leaders belong to $G$, or (ii) there is an equilibrium where $i$ has not activated their own pool and has delegated their whole stake to pools whose pool leaders belong to $G$, which have the highest desirability.

   We will prove that in both cases $i$'s non-myopic utility will not be higher than their current lower utility of the outer game denoted by $u_i^{\text{outer,low}}(\vec{m}^*, \vec{\lambda}^*)$.

   As a result, $u_i^{\text{outer,up}}(m_i, \vec{m}_{-i}^*, \lambda_i, \vec{\lambda}_{-i}^*) \leq u_i^{\text{outer,low}}(\vec{m}^*, \vec{\lambda}^*)$.

   In more detail:

   – In case (ii), by Lemma 5, their utility is at most
   
   $$(P(s_{k+1},c_{k+1}) + \epsilon'\cdot(1-\alpha))\cdot\frac{s_i}{\beta}$$
   $$< u_i^{\text{outer,low}}(\vec{m}^*, \vec{\lambda}^*).$$

– In case (i), their utility is

$$[m_i + (1 - m_i) \cdot \frac{\lambda_i}{\lambda_i}] \cdot (r(\lambda_i, \lambda_i) - c_i)$$

$$+ (P(s_{k+1}, c_{k+1}) + \epsilon' \cdot (1 - \alpha)) \cdot \frac{s_i - \lambda_i}{\beta}$$

$$\leq P(\lambda_i, c_i) \cdot (0 + (1 - 0) \cdot \frac{\lambda_i}{\beta}) +$$

$$(P(s_{k+1}, c_{k+1}) + \epsilon' \cdot (1 - \alpha)) \cdot \frac{s_i - \lambda_i}{\beta}$$

$$\overset{P(\lambda_i, c_i) \leq P(s_i, c_i), m_i^* > 0}{<} u_i^{\text{outer,low}}(\vec{m}^*, \vec{\lambda}^*).$$

We can prove that *there is no equilibrium in the inner game determined by $(m_i, \vec{m}^*_{-i}, \lambda_i, \vec{\lambda}^*_{-i})$, where the players in G other than i have not activated their own pools*, in the same way as the first case of the proof of Lemma 7. Note that also in this inner game determined by $(m_i, \vec{m}^*_{-i}, \lambda_i, \vec{\lambda}^*_{-i})$: (i) When these players activate their own pools, then these pools always have rank less or equal to $k$, regardless which other pools have been activated, and (ii) no other pool has strictly higher desirability and offers these players higher utility as members than they get by running their own pools.

Now we will prove by contradiction that *there is no equilibrium in the inner game defined by $(m_i, \vec{m}^*_{-i}, \lambda_i, \vec{\lambda}^*_{-i})$, where the $(k+1)$-st player does not activate their own pool*.

Let us suppose that there is a joint strategy $\vec{S}^{(m_i, \vec{m}^*_{-i}, \lambda_i, \vec{\lambda}^*_{-i})}$ that is an equilibrium of the inner game and for which player $k+1$ has not activated their own pool. Then by Lemma 5 it holds:

$$u_{k+1}(\vec{S}^{(m_i, \vec{m}^*_{-i}, \lambda_i, \vec{\lambda}^*_{-i})})$$
$$\leq (P(s_{k+1}, c_{k+1}) + \epsilon' \cdot (1 - \alpha)) \cdot \frac{s_{k+1}}{\beta}.$$

Then if they choose a different strategy ${S'}_{k+1}^{(m_i, \vec{m}^*_{-i}, \lambda_i, \vec{\lambda}^*_{-i})}$ where they activate their own pool, their utility can be increased:

$$u_{k+1}({S'}_{k+1}^{(m_i, \vec{m}^*_{-i}, \lambda_i, \vec{\lambda}^*_{-i})}, \vec{S}_{-(k+1)}^{(m_i, \vec{m}^*_{-i}, \lambda_i, \vec{\lambda}^*_{-i})})$$

$$= P(s_{k+1}, c_{k+1}) \cdot (1 - \frac{\epsilon' \cdot \alpha}{P(s_{k+1}, c_{k+1})}) \cdot \frac{s_{k+1}}{\beta}$$

$$+ \frac{\epsilon' \cdot \alpha}{P(s_{k+1}, c_{k+1})} \cdot P(s_{k+1}, c_{k+1})$$

$$= (P(s_{k+1}, c_{k+1}) + \epsilon' \cdot (1 - \alpha)) \cdot \frac{s_{k+1}}{\beta}$$

$$- \epsilon' \cdot \frac{s_{k+1}}{\beta} + \epsilon' \cdot \alpha$$

$$= u_{k+1}(\vec{S}^{(m_i, \vec{m}^*_{-i}, \lambda_i, \vec{\lambda}^*_{-i})}) - \epsilon' \cdot \frac{s_{k+1}}{\beta} + \epsilon' \cdot \alpha$$

$$\overset{\alpha > \frac{s_{k+1}}{\beta}}{>} u_{k+1}(\vec{S}^{(m_i, \vec{m}^*_{-i}, \lambda_i, \vec{\lambda}^*_{-i})}).$$

The intuition behind this is that $k+1$ can activate their own pool, that belongs to the $k$ most desirable pools, and in this way can increase their utility, because of the margin that they will take. Note that the desirability of their pool is worse than the desirability of pools whose pool leaders are players in $G$, but the difference is small, and thus being pool leader is more profitable for $k+1$ than being member of one of their pools.

Recall that in the inner game determined by $(m_i, \vec{m}^*_{-i}, \lambda_i, \vec{\lambda}^*_{-i})$, if player $k+1$ activates their own pool, this pool belongs to the $k$ most desirable pools, because $P(s_{k+1}, c_{k+1}) - \epsilon' \cdot \alpha > P(s_{k+2}, c_{k+2})$, where $P(s_{k+1}, c_{k+1}) - \epsilon' \cdot \alpha$ is its desirability. Note that the desirability of the other pools $\notin G$ is at most $P(s_{k+2}, c_{k+2})$ and that the desirability of $\pi_i$, if it is activated, is also lower than $P(s_{k+1}, c_{k+1}) - \epsilon' \cdot \alpha$. So there exist at most $k-1$ pools activated with higher desirability than that of the $(k+1)$-st player, which causes player $(k+1)$'s pool to belong to the $k$ most desirable pools when it is activated.

- No player $i \in G$ has an incentive to make their margin smaller than $m_i^*$, given that their pool already belongs to the best $k$ pools in all the equilibria of the inner game determined by $(\vec{m}^*, \vec{\lambda}^*)$.

    In more detail: If an $i \in G$ decreases their margin by choosing $m_i < m_i^*$ and chooses an arbitrary $\lambda_i \leq s_i$, then in the best case there is an equilibrium of the inner game where $\pi_i$ will again belong to the $k$ most desirable pools, and as a result:

    $$u_i^{\text{outer,up}}(m_i, \vec{m}^*_{-i}, \lambda_i, \vec{\lambda}^*_{-i})$$
    $$\leq (m_i + (1 - m_i) \cdot \frac{\lambda_i}{\beta}) \cdot P(s_i, c_i)$$
    $$+ \frac{s_i - \lambda_i}{\beta} \cdot (P(s_{k+1}, c_{k+1}) + \epsilon' \cdot (1 - \alpha))$$
    $$\stackrel{m_i < m_i^*}{<} (m_i^* + (1 - m_i^*) \cdot \frac{\lambda_i}{\beta}) \cdot P(s_i, c_i)$$
    $$+ \frac{s_i - \lambda_i}{\beta} \cdot (P(s_{k+1}, c_{k+1}) + \epsilon' \cdot (1 - \alpha))$$
    $$\leq u_i^{\text{outer,low}}(\vec{m}^*, \vec{\lambda}^*).$$

- No player $i \in G$ has incentives to commit less stake to their pool in the outer game, because the other existing pools have the same desirability and will not give it higher utility. In more detail: If a player $i \in G$ chooses margin equal to $m_i^*$, but $\lambda_i < s_i = \lambda_i^*$, then in the best case their pool will belong to the $k$ most desirable pools, and using Lemma 5, we will have:

    $$u_i^{\text{outer,up}}(\vec{m}^*, \lambda_i, \vec{\lambda}^*_{-i})$$
    $$\leq (m_i^* + (1 - m_i^*) \cdot \frac{\lambda_i}{\beta}) \cdot P(s_i, c_i)$$
    $$+ \frac{s_i - \lambda_i}{\beta}(P(s_{k+1}, c_{k+1}) + \epsilon' \cdot (1 - \alpha))$$
    $$\leq u_i^{\text{outer,low}}(\vec{m}^*, \vec{\lambda}^*).$$

- Player $i \in G$ has no incentives to set $\lambda_i = 0$, because using Lemma 5, we can prove that in any equilibrium of the inner game determined by $(m_i, \vec{m}^*_{-i}, 0, \vec{\lambda}^*_{-i})$, where they have not activated their own pool, as $\lambda_i = 0$, their utility for being a pool member of other pools will be lower than $u_i^{\text{outer,low}}(\vec{m}^*, \vec{\lambda}^*)$.

- Player $i \notin G$ has no incentives to choose margin $m_i \neq m_i^*$ or to commit stake $\lambda_i$ less than $s_i$, because we can prove in the same way as in the second case of the proof of Lemma 7 that in the inner game determined by $(m_i, \vec{m}^*_{-i}, \lambda_i, \vec{\lambda}^*_{-i})$, there is no equilibrium where they have activated their own pool, so margin and stake committed to their own pool do not have an impact on their utility, which by Lemma 5 will be at most $\frac{s_i}{\beta} \cdot (P(s_{k+1}, c_{k+1}) + \epsilon' \cdot (1 - \alpha))$. For this proof, it is important that players $\notin G$ cannot lower the desirability of the pools activated by players in $G$ by allocating stake to them strategically, because desirability does not depend on pool size. In addition to that, even if player $k+1$ sets margin zero, the desirability of their pool remains strictly lower than the desirability of the pools of all the players of $G$.